\documentclass[a4paper,fleqn]{cas-sc}

\usepackage[authoryear]{natbib}

\def\tsc#1{\csdef{#1}{\textsc{\lowercase{#1}}\xspace}}
\tsc{WGM}
\tsc{QE}

\usepackage{lineno}
\modulolinenumbers[5]

\usepackage{mathtools}

\usepackage{algorithm}
\usepackage{algpseudocode}

\usepackage{tabularx}
\usepackage{adjustbox}
\usepackage{longtable}
\usepackage{ragged2e}
\usepackage{enumitem}

\usepackage{rotating}
\usepackage{pdflscape}

\usepackage{tikz}
\usetikzlibrary{
  arrows.meta,
  shadows,
  shapes.geometric
}

\usepackage[edges]{forest}

\usepackage{tcolorbox}
\tcbuselibrary{breakable}

\usepackage{svg}

\usepackage{siunitx}

\newcommand{\trans}{\ensuremath{T}}
\newcommand{\tran}{\ensuremath{t}}

\newcommand{\events}{\ensuremath{\Sigma}}
\newcommand{\event}{\ensuremath{\sigma}}

\newcommand{\user}{\ensuremath{u}}
\newcommand{\users}{\ensuremath{U}}

\newcommand{\cons}{\ensuremath{R}}

\newcommand{\tokens}{\ensuremath{N}}

\newcommand{\cluster}{\ensuremath{C}}

\newcommand{\emb}{\ensuremath{z}}

\newcommand{\method}{BUBA\xspace}

\newcolumntype{Y}[1]{%
  >{\RaggedRight\arraybackslash}p{#1}%
}

\tcbset{
  rqbox/.style={
    colframe=gray!75!black,
    colback=gray!5!white,
    fonttitle=\bfseries,
    boxrule=0.6pt,
    arc=2pt,
    left=6pt,
    right=6pt,
    top=5pt,
    bottom=5pt,
    before skip=6pt,
    after skip=6pt,
    breakable
  }
}

\newcommand{\yesmark}{%
  \tikz[baseline=-0.55ex]
    \fill (0,0) circle[radius=2.2pt];%
}

\newcommand{\nomark}{%
  \tikz[baseline=-0.55ex]
    \draw (0,0) circle[radius=2.2pt];%
}

\newcommand{\partmark}{%
  \tikz[baseline=-0.55ex]{%
    \begin{scope}
      \clip (0,0) circle[radius=2.2pt];
      \fill (-2.2pt,-2.2pt) rectangle (0pt,2.2pt);
    \end{scope}
    \draw (0,0) circle[radius=2.2pt];
  }%
}

\begin{document}
\let\WriteBookmarks\relax
\def\floatpagepagefraction{1}
\def\textpagefraction{.001}

\shorttitle{Persistent Behavioural Patterns for Blockchain Forensics}    

\shortauthors{Zelenyanszki et al.}  

\title [mode = title]{Discovering Persistent Behavioural Patterns for Interpretable Blockchain Forensics}  



%

\author[1]{Dorottya Zelenyanszki}[orcid=0009-0004-1300-0357]

\cormark[1]


\ead{dora.zelenyanszki@griffithuni.edu.au}



\affiliation[1]{organization={Griffith University},
            addressline={170 Kessels Road}, 
            city={Nathan},
            postcode={4111}, 
            state={QLD},
            country={Australia}}

\author[1]{Zh\'e H\'ou}[orcid=0000-0001-7164-0580]


\ead{z.hou@griffith.edu.au}



\author[2]{Kamanashis Biswas}[orcid=0000-0003-3719-8607]



\ead{Kamanashis.Biswas@acu.edu.au}



\affiliation[2]{organization={Australian Catholic University},
            addressline={1100 Nudgee Road}, 
            city={Banyo},
            postcode={4014}, 
            state={QLD},
            country={Australia}}

\author[1]{Vallipuram Muthukkumarasamy}[orcid=0000-0002-6787-6379]



\ead{v.muthu@griffith.edu.au}




\cortext[1]{Corresponding author}



\begin{abstract}
Public blockchain data enables large-scale DeFi-related analysis, but many existing approaches are application-specific, difficult to scale, or hard to interpret. This research proposes a scalable, application-agnostic framework for \emph{persistent behavioural pattern discovery} from large-scale blockchain activity. It constructs behaviour sentences enriched with contract, token and market context, then applies a two-step embedding process: sentence-level embeddings capture individual actions, while sequence-level embeddings capture user behaviour over time. An interpretable behavioural profiler characterizes discovered communities through behavioural motifs, routines, temporal dynamics, entity exposure, and suspiciousness evidence. Evaluation on Ethereum using over 30 million transactions shows that the framework uncovers both routine and malicious behavioural patterns, including decentralised exchange (DEX) trading, NFT activity, phishing, bot operations, oracle manipulation, and rug-pull schemes. Importantly, many patterns remain stable across independent observation windows, enabling the identification of long-term behaviours beyond a single analysis period. The proposed framework combines scalability, interpretability, and persistence analysis, supporting blockchain forensic investigation, behavioural attribution, and threat discovery.
\end{abstract}




\maketitle

\section{Introduction}
\label{sec:introduction}
DeFi has rapidly evolved into a large ecosystem of smart contracts, tokens, liquidity pools, and user-driven protocol interactions. Its security challenges extend beyond smart-contract correctness to protocol design, market structure, and the economic incentives governing participant behaviour~\citep{10.1145/3558535.3559780}. At the implementation level, mechanisms such as upgradeable smart contracts can introduce framework-specific faults and complicate the testing of upgrade paths, potentially resulting in deployed vulnerabilities and security incidents~\citep{11173264}. The operational consequences remain substantial. For example, the De.Fi REKT report for October 2025\footnote{\url{https://tinyurl.com/t2e294nz}} documented losses of \$38.63 million across nine exploits in a single month. Detecting and understanding such risks requires analysing blockchain data that is publicly accessible but heterogeneous, massive in volume, and inherently temporal, making it both a challenging and promising setting for machine learning~\citep{10.1145/3728474}. Public availability, however, does not make on-chain behaviour immediately interpretable. Transactions, logs, contract calls, token movements, and event arguments provide low-level records of activity, but do not directly reveal user roles, recurring routines, or persistent behavioural patterns. A central challenge is therefore to transform large volumes of heterogeneous on-chain activity into representations that support population-level behavioural analysis, rather than limiting analysis to individual transactions or incidents.

Existing work has advanced blockchain security and DeFi risk analysis in several directions. These include forensic tracing and suspicious-flow reconstruction~\citep{lin2025risktaggerllmbasedagentautomatic,11198865}, behavioural representation learning and user classification~\citep{10.1145/3690624.3709322,bonifazi2022defining}, and attack- or scam-specific detection~\citep{10.1145/3696410.3714682,10740038, zhong2025detectingvariousdefiprice,236240}. These approaches are valuable, and, in several cases, highly scalable within their own problem settings, but they are often transaction- or flow-centric, attack-specific, label-dependent, tied to particular protocols, or focused on individual suspicious entities. They therefore provide limited support for discovering broad, persistent, and interpretable behavioural structure across large user populations, especially when suspicious evidence is sparse, mixed with ordinary activity, and evolves over time. In particular, they do not directly address deriving long-term, persistent behavioural patterns from millions of transactions while jointly incorporating direct suspicious labels and exposure-based suspicious evidence. They also provide limited support for explaining how users organise into recurring behavioural groups, how suspicious evidence is embedded within ordinary activity, and whether these behavioural patterns persist across independent and longer observation windows.

This paper addresses these gaps through a scalable, application-agnostic framework for \emph{persistent behavioural pattern discovery} of Ethereum users. It transforms raw transaction activity into behavioural sentences, embeds them through a two-step representation process, clusters users according to their sequence-level behaviour, and profiles the resulting groups using behavioural and suspiciousness-oriented evidence. The framework operates on user histories, preserves sequential activity, and does not require labels for malicious or benign, but to reveal how ordinary and suspicious behaviours are organised across the population and whether these patterns persist over time. It also produces interpretable cluster-level and tag-level profiles while remaining practical at million-user scale.

The main contributions of this paper are threefold:
\begin{itemize}

\item We propose a scalable, application-agnostic framework for \emph{persistent behavioural pattern discovery} that converts blockchain transactions and decoded logs into behavioural sentences preserving action order, event semantics, entity types, and market signals learns user-level sequence representations, clusters users without labels, and profiles the resulting behavioural groups.

\item We establish the most suitable framework setting through a systematic, multi-stage evaluation of sentence embedding models, sequence aggregation methods, training objectives, sequence-length strategies, clustering algorithms and resolutions, and behavioural semantics, balancing representation quality, clustering structure and stability, scalability, length leakage, and the concentration of pre-labelled behaviours.

\item We develop an interpretable behavioural profiler that explains discovered clusters through behavioural motifs, ordered routines, activity and temporal patterns, diversity, concentration, entity exposure, and separated direct-label and exposure-based suspiciousness evidence, enabling the identification of persistent ordinary and suspicious behavioural patterns across independent monthly and cumulative multi-month windows.

\end{itemize}

The remainder of the paper is organised as follows. Section~\ref{sec:related_work} reviews the relevant literature, and Section~\ref{sec:framework} presents the proposed framework. Section~\ref{sec:experiments} describes the experiments. Section~\ref{sec:analysis} examines the resulting behavioural profiles and their persistence across observation windows. Sections~\ref{sec:limits} and \ref{sec:conclusion} discuss the limitations and conclude the paper.
\section{Related work}
\label{sec:related_work}
Relevant research on blockchain security and DeFi risk includes forensic tracing and analysis, behavioural representation learning, attack-specific detection, pre-incident attack inference, and governance- or ecosystem-level analysis. Although these directions have improved visibility into unsafe activity, they often focus on specific attacks, contracts, transactions, labelled classification tasks, or curated cases, limiting their ability to recover unsupervised, interpretable, and long-term behavioural patterns across large user populations.

Forensic tracing systems focus on reconstructing suspicious transaction flows. \citet{lin2025risktaggerllmbasedagentautomatic} propose RiskTagger, an LLM-guided agent that extracts incident clues, traces laundering-related fund flows through iterative reasoning over on-chain evidence, and generates evidence-grounded forensic reports. \citet{11198865} present ABCTRACER, which combines semantic log extraction, named entity recognition, and information retrieval to trace suspicious cross-chain bridge flows. These systems demonstrate the value of semantic recovery and reasoning for identifying and explaining suspicious flows, but do not aim to recover stable behavioural organisation across large user populations.

Behavioural representation learning is more closely related to our setting. \citet{10.1145/3690624.3709322} introduce Chainlet Orbits, which captures structural roles in temporal blockchain graphs to produce scalable and interpretable address embeddings. \citet{bonifazi2022defining} represent Ethereum users through multivariate temporal profiles for multi-class classification. \citet{tovanich2023fingerprinting} learn Bitcoin taint-flow fingerprints by extracting taint-flow graphs, sampling random walks, and embedding money-flow patterns for classification and clustering. These studies demonstrate that blockchain entities can be represented through structural, temporal, and flow-based behaviour, but they mainly support address classification, Bitcoin money-flow analysis, or clustering without profiler-based interpretation. In contrast, our framework models Ethereum user histories using smart-contract calls, decoded events, token standards, approvals, swaps, NFT movements, and protocol interactions, then interprets the resulting clusters using post hoc suspiciousness evidence and cross-window persistence analysis.

Vulnerability and attack-specific detection methods are effective within predefined threat classes. \citet{10.1145/3696410.3714682} formalise attack behaviour using temporal logic for runtime monitoring in MoE. \citet{10740038} propose DeFiGuard, which detects price manipulation from transaction-derived cash-flow graphs, whereas \citet{zhong2025detectingvariousdefiprice} use large language models to recover DeFi operations and detect price manipulation. \citet{IBBA2025100759} combine software metrics and topic modelling with machine-learning classifiers to detect vulnerabilities in Ethereum smart contracts, reporting \(\num{97.7}\%\) accuracy for binary classification and an improved F1-score of \(\num{0.881}\) after incorporating topic information. \citet{11201921} use temporal graph neural networks to detect rug-pull risks, with memory and attention mechanisms used to model dynamic token-transaction patterns, while \citet{236240} use symbolic execution to detect honeypot contracts. These methods are effective for detecting known vulnerabilities, attacks, and scam mechanisms within well-defined settings, but their reliance on specialised threat definitions, labelled data, or contract-level analysis makes them less suitable for broad behavioural discovery across large user populations, particularly when the aim is to identify latent suspicious groups without predefining the behaviour of interest.

Pre-incident inference and ecosystem-level studies address other dimensions of blockchain risk. \citet{10.1145/3691620.3695526} analyse adversarial contracts before exploit transactions to infer likely attack intentions, but their approach remains contract- and incident-centric. Although this supports operational defence before an exploit occurs, it does not target stable user-level behavioural organisation. Broader studies examine DAO governance weaknesses~\citep{10891888}, rug-pull causes, datasets, and detection tools~\citep{sun2024sokcomprehensiveanalysisrug}, and longitudinal token and liquidity-pool activity involving token spammers, disposable-token rug pulls, and sniper bots~\citep{287184}. The governance analysis shows that risk can arise from both technical and procedural failures, while the rug-pull study highlights the limited coverage of existing datasets and tools. The longitudinal ecosystem analysis further identifies token spammers, disposable-token rug pulls, and sniper bots as recurring malicious roles. These studies broaden blockchain risk analysis by examining governance processes, rug-pull causes and dataset coverage, and longitudinal token and liquidity-pool ecosystems. However, each addresses a particular aspect of the ecosystem rather than providing a general framework for discovering and interpreting behavioural patterns across large user populations.

Overall, existing work provides strong foundations in forensic tracing, behavioural representation, vulnerability and attack detection, and ecosystem analysis. However, these lines of work do not jointly support large-scale user-history representation, sequence-aware modelling, unsupervised grouping, post hoc interpretation using suspiciousness evidence, and interpretable behavioural profiling. Our framework addresses this gap by combining sentence-based behavioural abstraction, sequence-level embedding, clustering, and profiling. Table~\ref{tab:related_work_positioning} summarises this positioning.

\begin{table}[pos=!t]
\centering
\caption{Positioning of the related work. \emph{Hist., Seq., Unsup., Agn.,
Prof., and LT denote user-history modelling, sequence-aware modelling,
unsupervised discovery, application-agnostic analysis, interpretable profiling,
and long-term persistence, respectively. Filled, half-filled, and unfilled
circles denote yes, partial, and no.}}
\label{tab:related_work_positioning}
\small
\setlength{\tabcolsep}{2pt}
\renewcommand{\arraystretch}{1.08}
\begin{tabular}{@{}p{0.46\columnwidth}cccccc@{}}
\hline
\textbf{Related work} &
\textbf{Hist.} &
\textbf{Seq.} &
\textbf{Unsup.} &
\textbf{Agn.} &
\textbf{Prof.} &
\textbf{LT} \\
\hline

Forensic tracing and analysis
& \partmark & \partmark & \nomark & \partmark & \partmark & \nomark \\

Bitcoin taint-flow fingerprints
& \yesmark & \partmark & \partmark & \partmark & \nomark & \nomark \\

Representation and classification
& \yesmark & \partmark & \nomark & \partmark & \partmark & \partmark \\

Vulnerability and attack detection
& \partmark & \partmark & \nomark & \nomark & \partmark & \nomark \\

Governance and ecosystem studies
& \partmark & \partmark & \partmark & \nomark & \yesmark & \partmark \\
\hline

\textbf{Proposed work}
& \yesmark & \yesmark & \yesmark & \yesmark & \yesmark & \yesmark \\
\hline
\end{tabular}
\end{table}
\section{The proposed framework}
\label{sec:framework}
\begin{figure*}[pos=!t]
\centering
\includegraphics[width=0.8\textwidth]{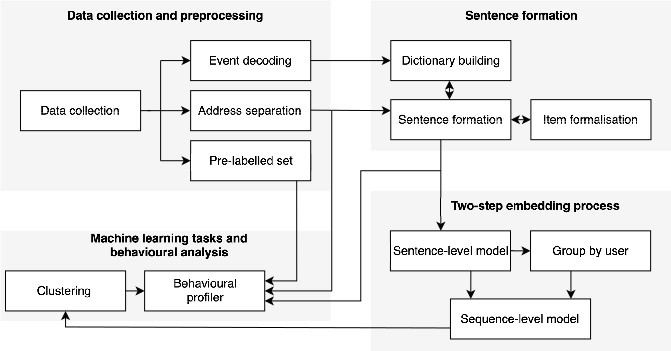}
\caption{Overview of the proposed framework.}
\label{fig:framework}
\end{figure*}

This section presents the proposed framework. It abstracts transactions and logs into behavioural sentences, embeds them into user-level representations, and clusters users without labels. The resulting groups are then profiled using shared routines, entity exposure, temporal patterns, and post hoc suspiciousness evidence. The framework is visualised in Fig.~\ref{fig:framework}.

\subsection{Data Collection and Preprocessing}
Ethereum transaction data covering blocks 16250000 to 16749999 was collected from the XBlock-ETH dataset~\cite{Zheng2020}. For each transaction, we extract the transaction hash, sender and receiver addresses, timestamp, value, gas usage, and decoded event-log information, including event emitters, event names, method IDs, and typed arguments. Event decoding is performed through Infura API queries.

Collected transactions are examined to extract candidate addresses from:
(i) \texttt{from}/\texttt{to} fields,
(ii) decoded log arguments with typed address values (dictionaries of the form \(\{\texttt{type}=\texttt{address}, \texttt{value}=\texttt{0x...}\}\)), avoiding arbitrary string mining,
(iii) event emitter addresses, and
(iv) created-contract fields (e.g., \texttt{contractAddress}, \texttt{createdContractAddress}, and \texttt{toCreate}).
Using these signals together with XBlock-ETH and the Zellic contracts dataset\footnote{\url{https://huggingface.co/datasets/Zellic/all-ethereum-contracts}}, addresses are separated into contract-like entities versus EOA candidates, and contract candidates are further probed for ERC-20/721 traits via standard \texttt{eth\_call} selectors, including \texttt{name},
\texttt{symbol}, \texttt{decimals}, \texttt{totalSupply},
\texttt{supportsInterface} for ERC-721 and metadata IDs, and
\texttt{balanceOf(0x0)}. Some EOAs are promoted to potential users without further checks, including observed transaction senders and creator addresses labelled as EOAs in XBlock-ETH. Remaining EOA candidates are filtered by deterministic static heuristics for zero, dead, precompile, \texttt{0xeeee...}, near-zero-\texttt{prefix}, and small-numeric addresses, for example addresses below \(2^{96}\). For the remaining unresolved addresses, we query \texttt{eth\_getTransactionCount} at the latest available state: addresses with nonce \(\geq \num{1}\) are added to the user set, while addresses with nonce \(\num{0}\) are retained in the filtered set. The first iteration on the first-month window yielded \num{6045497} users, \num{3117292} contracts, \num{1343354} ERC-20 tokens, \num{259062} ERC-721 tokens, with \num{1317947} addresses filtered out.

Token-level context is added where available. ERC-20 information comes from CoinGecko and ERC-721 information from OpenSea, providing metadata such as supply, price, volume, ownership, sales, and floor price. Addresses are mapped to readable user and contract IDs. Transactions are then linked to users: sent transactions are \emph{primary}, while event-linked transactions are \emph{secondary}. The resulting histories, address sets, and market data form the input to sentence formation.

External suspiciousness labels are collected from community and research sources, including the De.Fi REKT database\footnote{\url{https://de.fi/rekt-database/}}, DefiLlama\footnote{\url{https://defillama.com/hacks}}, DeFiHackLabs\footnote{\url{https://github.com/SunWeb3Sec/DeFiHackLabs/tree/main}}, ImmuneBytes reports\footnote{\url{https://tinyurl.com/68796ant}}, and the bot-address dataset of~\cite{10.1145/3589335.3651959}. The final pre-labelled set contains \num{358} unique addresses (some resources present overlaps), with category-level assignments including \num{68} phishing users, \num{50} phishing contracts, \num{94} malicious contracts, \num{96} exploiters, and \num{58} exploited contracts. These labels are not used to train, guide, or constrain clustering. They are introduced only after clustering and profiling as high-confidence external evidence.

\subsection{Sentence Formation}
Sentence formation converts transactions and decoded logs into compact behavioural text while preserving action order, event semantics, entity types, and market signals. Table~\ref{tab:sentence_examples} shows examples of final behavioural sentences. Let,

\[
\mathbf{\trans}
=
\{\mathbf{\tran}_{1},\mathbf{\tran}_{2},\ldots,\mathbf{\tran}_{k}\}
\]
be the transaction set. Each transaction \(\mathbf{\tran}_{i}\) contains an ordered decoded-event list
\[
\events(\mathbf{\tran}_{i})
=
[\event_{i,1},\event_{i,2},\ldots,\event_{i,l_i}],
\]
where \(\event_{i,j}\), with \(1\leq j \leq l_i\), represents a decoded event object, and \(l_i\) is the number of decoded events in \(\mathbf{\tran}_{i}\).

An event dictionary maps canonical event signatures to stable aliases. For example, \texttt{Transfer(address,address\\,uint256)} may be mapped to \texttt{transfer1}. Let \(\phi(\event_{i,j})\) denote the canonical signature of event \(\event_{i,j}\), and let \(\psi(\phi(\event_{i,j}))\) denote its alias.

For each transaction, we first construct an intermediate sentence \(\tilde{s}_i\) from transaction-level metadata and ordered event-level records. Each event record contains the event alias and the addresses appearing as event emitters or typed address arguments. These addresses are grouped using the address classification from the preprocessing stage: ERC-20 tokens, ERC-721 tokens, contracts, filtered addresses, and users.

The final sentence \(s_i\) is obtained by applying the transformation \(\tau(\tilde{s}_i)=s_i\). This transformation normalises transaction-level attributes, including value, gas usage, transaction type, and timestamp, into compact tokens. It also represents events by aliases, compresses repeated events with the same structure, includes token and market information on the first token occurrence within a transaction, and replaces contract and user mentions with compact identifiers. Applying \(\tau\) to all intermediate sentences gives the sentence set
\[
\{s_1,s_2,\ldots,s_k\}.
\]

\begin{table}[!t]
\centering
\caption{Typical samples of final behavioural sentences.}
\label{tab:sentence_examples}
\small
\setlength{\tabcolsep}{2.5pt}
\renewcommand{\arraystretch}{1.12}
\begin{tabular}{@{}
>{\raggedright\arraybackslash}p{0.24\columnwidth}
>{\raggedright\arraybackslash}p{0.70\columnwidth}
@{}}
\hline
\textbf{Example} &
\textbf{Final behavioural sentence} \\
\hline

No decoded events &
{\footnotesize\ttfamily 0 46.27k primary 1671841535} \\[2pt]

\rowcolor{gray!16}
Primary event-rich transaction &
{\footnotesize\ttfamily 0 104.43k primary 1671830579 transfer1 ERC20 UNI-V2 0 no market value C5352 repeats 2 sync1 C5352 swap1 C5352} \\[2pt]

Secondary event-rich transaction &
{\footnotesize\ttfamily 0.045 332.06k secondary 1671918791 transfer1 C1013179 A320147 approvalforall1 C1013179 C368034 transfer1 C1013179 A1305885 repeats 2 transfer1 C1013179 C708 A1305885 transfer1 C1013179 A1305885 self withdrawal1 C1013179} \\
\hline
\end{tabular}
\end{table}

Let \(\mathbf{\users}\) denote the set of users. For each user \(\mathbf{\user}_{r}\in\mathbf{\users}\), let
\[
Q(\mathbf{\user}_{r})
=
[\mathbf{\tran}_{r,1},\mathbf{\tran}_{r,2},\ldots,\mathbf{\tran}_{r,t_r}]
\]
be the time-ordered transaction sequence linked to that user. The corresponding behavioural sentence sequence is
\[
\mathcal{T}(\mathbf{\user}_{r})
=
[s_{r,1},s_{r,2},\ldots,s_{r,t_r}],
\]
where \(s_{r,j}\) is derived from transaction \(\mathbf{\tran}_{r,j}\). This sequence is the input to the two-step embedding process. Let,
\[
\mathbf{\users}^{+}
=
\{\mathbf{\user}_{r}\in\mathbf{\users}
\mid |\mathcal{T}(\mathbf{\user}_{r})|\geq \ell_{\min}\}
\]
denote the users, who are retained after applying the minimum behavioural-history threshold \(\ell_{\min}\).

\subsection{Two-Step Embedding Process}
The two-step embedding process maps each retained user's behavioural sentence sequence to a compact user-level representation. 
Rather than representing transactions independently, the framework embeds behavioural sentences and aggregates their ordered sequence to capture user-level routines.

In the first step, a pretrained sentence model maps each sentence to a fixed-length vector. Let
\[
g_\alpha : \mathcal{D}_{\mathrm{sent}} \rightarrow \mathbb{R}^{d_s}
\]
be the sentence embedding function for model \(\alpha\), where \(\mathcal{D}_{\mathrm{sent}}\) is the sentence domain and \(d_s\) is the embedding dimension. The sentence embedding sequence for user \(\mathbf{\user}_{r}\) is
\[
\mathbf{Z}^{\mathrm{sent}}(\mathbf{\user}_{r})
=
[\mathbf{z}^{\mathrm{sent}}_{r,1}, \mathbf{z}^{\mathrm{sent}}_{r,2},
\ldots, \mathbf{z}^{\mathrm{sent}}_{r,t_r}]
=
[g_\alpha(s_{r,1}), g_\alpha(s_{r,2}), \ldots, g_\alpha(s_{r,t_r})].
\]

In the second step, the sentence embedding sequence is converted into a single user embedding. Since user histories may have different lengths, sequences longer than the chosen maximum length \(L\) are cropped according to the selected policy, while shorter sequences are padded and masked during batching. Let
\[
\widehat{\mathbf{Z}}^{\mathrm{sent}}(\mathbf{\user}_{r})
=
[\widehat{\mathbf{z}}^{\mathrm{sent}}_{r,1},
\widehat{\mathbf{z}}^{\mathrm{sent}}_{r,2},
\ldots,
\widehat{\mathbf{z}}^{\mathrm{sent}}_{r,\hat{t}_r}]
\]
denote the processed sequence, where \(\hat{t}_r \leq \min(t_r,L)\). A sequence aggregation model then maps this sequence to a user-level vector. Let
\[
h_\beta : (\mathbb{R}^{d_s})^{\hat{t}_r} \rightarrow \mathbb{R}^{d_s}
\]
be the aggregation function for sequence model \(\beta\). The final embedding of user \(\mathbf{\user}_{r}\) is
\[
\mathbf{\emb}_{r}
=
h_\beta(\widehat{\mathbf{Z}}^{\mathrm{sent}}(\mathbf{\user}_{r})).
\]

Applying this process to all retained users yields
\[
\mathcal{E}
=
\{\mathbf{\emb}_1,\mathbf{\emb}_2,\ldots,\mathbf{\emb}_m\},
\]
where \(m=|\mathbf{\users}^{+}|\). Each vector represents one retained user's behavioural history and is used for downstream clustering.

\subsection{Clustering}
\label{subsec:clustering}
Given the user embedding set \(\mathcal{E}\), the clustering stage assigns retained users to behavioural groups. Formally, the objective is to learn a mapping
\[
f:
\mathbf{\users}^{+}
\rightarrow
\{\cluster_1,\cluster_2,\ldots,\cluster_n\}.
\]
where \(\cluster_1,\ldots,\cluster_n\) denote the discovered behavioural clusters. Users with similar learned behavioural representations are clustered together, revealing recurring retained-user patterns. The stage is fully unsupervised: pre-labelled addresses and external references are not used to train or constrain the clustering; they are incorporated only afterwards during profiling.

\subsection{Behavioural Profiler}
\label{subsec:profiler}
After clustering, the framework applies a behavioural profiler to describe the discovered groups. The profiler combines cluster membership with behavioural artefacts produced by the earlier stages, including intermediate and final behavioural sentences and the associated contract and token interactions. From these inputs, it derives motif abstractions, ordered motif sequences, activity and temporal profiles, entity-exposure profiles, and contract and token diversity or concentration patterns. This allows each cluster to be interpreted not only as a geometric grouping in the embedding space, but also as a behavioural group with shared actions, routines, and interaction patterns.

Available risk evidence, including pre-labelled addresses, risky entity references, and exploit-transaction references, is introduced only during profiling. It is used to interpret and evaluate groups after clustering rather than to learn user representations or determine cluster membership.

\paragraph{Behavioural base.}
The behavioural base brings together the information needed to describe each user's activity. For a user \(\mathbf{\user}_{r}\in\mathbf{\users}\), the profiler collects the corresponding intermediate and final behavioural sentence sequences, together with the contracts and tokens involved in those behaviours. We represent this user-level information as
\[
p_r
=
\bigl(
\mathbf{\user}_{r},
\widetilde{\mathcal{T}}(\mathbf{\user}_{r}),
\mathcal{T}(\mathbf{\user}_{r}),
\mathbf{\cons}_{r},
\mathbf{\tokens}_{r}
\bigr),
\]
where \(\widetilde{\mathcal{T}}(\mathbf{\user}_{r})\) and \(\mathcal{T}(\mathbf{\user}_{r})\) denote the intermediate and final behavioural sentence sequences, and \(\mathbf{\cons}_{r}\subseteq\mathbf{\cons}\) and \(\mathbf{\tokens}_{r}\subseteq\mathbf{\tokens}\) denote the associated contracts and tokens.

Although the implementation organises the underlying information as transaction-level records, \(p_r\) denotes the combined user-level profiler record for \(\mathbf{\user}_{r}\). Throughout this sub-section, the subscript \(r\) therefore refers to the same user and to the profiling artefacts derived for that user.

The complete profiler set is
\[
\mathcal{D}_{\mathrm{prof}}
=
\left\{
p_r
\mid
\mathbf{\user}_{r}\in\mathbf{\users}
\right\}.
\]

After applying the global minimum behavioural-history threshold, the retained profiler set is
\[
\mathcal{D}_{\mathrm{prof}}^{+}
=
\left\{
p_r
\mid
\mathbf{\user}_{r}\in\mathbf{\users}^{+}
\right\}.
\]

Thus, \(\mathcal{D}_{\mathrm{prof}}^{+}\) contains the profiler records of the users retained for subsequent behavioural analysis.

\paragraph{Ordered motif artefacts.}
For each retained user, the profiler extracts motifs from the behavioural sentence sequence. A motif is a compact behavioural unit preserving the main action and important entities involved, such as a token transfer, contract-mediated swap, approval, NFT transfer, or claim event. Let
\[
\mu:
\mathcal{D}_{\mathrm{sent}}
\rightarrow
\mathcal{M}^{*}
\]
denote the motif extraction function, where \(\mathcal{M}^{*}\) is the space of finite motif sequences. The ordered motif sequence of \(\mathbf{\user}_{r}\) is
\[
M(\mathbf{\user}_{r})
=
\mu(s_{r,1})
\,\Vert\,
\mu(s_{r,2})
\,\Vert\,
\ldots
\,\Vert\,
\mu(s_{r,t_r}),
\]
where \(\Vert\) denotes sequence concatenation. Its compact summary is
\[
B(\mathbf{\user}_{r})
=
\operatorname{Summary}
\bigl(
M(\mathbf{\user}_{r})
\bigr).
\]

The sequence \(M(\mathbf{\user}_{r})\) preserves behavioural order, while \(B(\mathbf{\user}_{r})\) provides a compact user-level summary for cluster-level comparison.

\paragraph{Cluster membership.}
The profiler links the retained behavioural records to the clustering result \(f\). The retained users assigned to cluster \(\cluster_j\) are
\[
\mathbf{\users}_{j}
=
\left\{
\mathbf{\user}_{r}\in\mathbf{\users}^{+}
\mid
f(\mathbf{\user}_{r})=\cluster_j
\right\}.
\]

\paragraph{User-level evidence.}
For each retained user \(\mathbf{\user}_{r}\), the profiler constructs complementary activity, entity-exposure, and direct risk-label profiles.

The activity profile of \(\mathbf{\user}_{r}\) is
\[
\pi^{\mathrm{act}}_r
\equiv
\pi^{\mathrm{act}}(\mathbf{\user}_{r})
=
\operatorname{ActProf}
\bigl(
p_r,
M(\mathbf{\user}_{r})
\bigr).
\]

The function \(\operatorname{ActProf}\) derives activity measurements from the sentence histories in \(p_r\) and the ordered motif sequence \(M(\mathbf{\user}_{r})\). The resulting profile records activity volume, the number of represented transactions, first and last activity times, active span, transaction frequency, inter-record gaps, and coarse temporal patterns.

The entity-exposure profile is
\[
\pi^{\mathrm{ent}}_r
\equiv
\pi^{\mathrm{ent}}(\mathbf{\user}_{r})
=
\operatorname{EntProf}
\bigl(
p_r,
\mathcal{L}_{\mathrm{ent}}
\bigr),
\]
where \(\mathcal{L}_{\mathrm{ent}}\) denotes the available risky entity references.

The function \(\operatorname{EntProf}\) analyses the contract and token interactions represented in \(p_r\). It records the distinct entities involved, counts repeated interactions, measures entity diversity and concentration, identifies frequently interacted entities, and compares them with the risky entity references in \(\mathcal{L}_{\mathrm{ent}}\).

The direct risk-label profile is
\[
\pi^{\mathrm{lab}}_r
\equiv
\pi^{\mathrm{lab}}(\mathbf{\user}_{r})
=
\operatorname{LabProf}
\bigl(
\mathbf{\user}_{r},
\mathcal{L}_{\mathrm{addr}}
\bigr),
\]
where \(\mathcal{L}_{\mathrm{addr}}\) is the pre-labelled address set.

The function \(\operatorname{LabProf}\) retrieves the entries in \(\mathcal{L}_{\mathrm{addr}}\) whose address is \(\mathbf{\user}_{r}\) and records the associated risk labels, if any. These labels are attached only after user representations and cluster memberships have been produced.

\paragraph{Cluster-level evidence.}
The profiler aggregates the retained user-level evidence within each cluster. 

The cluster risk profile and its risk-type breakdown are obtained by aggregating the direct risk-label and entity-exposure profiles of the cluster members:
\[
\bigl(
\Pi^{\mathrm{risk}}_j,
\Delta^{\mathrm{risk}}_j
\bigr)
=
\operatorname{RiskProf}
\left(
\left\{
\bigl(
\pi^{\mathrm{lab}}(\mathbf{\user}_{r}),
\pi^{\mathrm{ent}}(\mathbf{\user}_{r})
\bigr)
\mid
\mathbf{\user}_{r}\in\mathbf{\users}_{j}
\right\}
\right).
\]

For each cluster member, \(\operatorname{RiskProf}\) combines the user's direct risk-label and entity-exposure profiles and aggregates this evidence across \(\mathbf{\users}_{j}\). The resulting profile \(\Pi^{\mathrm{risk}}_j\) summarises the available cluster-level risk evidence, while \(\Delta^{\mathrm{risk}}_j\) provides its breakdown by risk type.

The exploit-transaction profile is
\[
\Pi^{\mathrm{tx}}_j
=
\operatorname{TxProf}
\left(
\mathbf{\users}_{j},
\left\{
p_r
\mid
\mathbf{\user}_{r}\in\mathbf{\users}_{j}
\right\},
\mathcal{L}_{\mathrm{tx}}
\right),
\]
where \(\mathcal{L}_{\mathrm{tx}}\) denotes the available exploit-transaction references.

The sentence histories stored in \(p_r\) preserve the source-transaction references associated with each sentence, allowing the represented behaviour to be traced back to the transactions from which it was derived. The function \(\operatorname{TxProf}\) matches these transaction references against \(\mathcal{L}_{\mathrm{tx}}\) and aggregates the resulting matches. The profile \(\Pi^{\mathrm{tx}}_j\) records the number and proportion of users with exploit-transaction exposure, the proportion of represented activity involving such transactions, and the concentration of the exposure across cluster members. Contact with an exploit-labelled transaction is treated as contextual evidence and does not by itself imply that the entire cluster is malicious.

The defining motifs of cluster \(\cluster_j\) are obtained by comparing the motif summaries of users inside and outside the cluster:
\[
D^{\mathrm{mot}}_j
=
\operatorname{MotifRank}
\left(
\left\{
B(\mathbf{\user}_{r})
\mid
\mathbf{\user}_{r}\in\mathbf{\users}_{j}
\right\},
\left\{
B(\mathbf{\user}_{r})
\mid
\mathbf{\user}_{r}
\in
\mathbf{\users}^{+}\setminus\mathbf{\users}_{j}
\right\}
\right).
\]

Similarly, the defining motif sequences are
\[
D^{\mathrm{seq}}_j
=
\operatorname{SeqRank}
\left(
\left\{
M(\mathbf{\user}_{r})
\mid
\mathbf{\user}_{r}\in\mathbf{\users}_{j}
\right\},
\left\{
M(\mathbf{\user}_{r})
\mid
\mathbf{\user}_{r}
\in
\mathbf{\users}^{+}\setminus\mathbf{\users}_{j}
\right\}
\right).
\]

The functions \(\operatorname{MotifRank}\) and \(\operatorname{SeqRank}\) compare the evidence observed among users inside \(\cluster_j\) with that observed among retained users outside the cluster. They rank motifs and motif sequences using their occurrence frequency, user coverage, and cluster-versus-rest distinctiveness. The resulting artefacts therefore describe both which behavioural units distinguish a cluster and how those units are ordered. Their semantic interpretation is
\[
\omega_j
=
\operatorname{Read}
\bigl(
D^{\mathrm{mot}}_j,
D^{\mathrm{seq}}_j
\bigr).
\]

The function \(\operatorname{Read}\) converts the ranked motif and sequence evidence into a human-understandable description of the cluster's principal actions, interacted entity types, and recurring ordered routines.

The cluster temporal and diversity profile is
\[
\Pi^{\mathrm{td}}_j
=
\operatorname{TDProf}
\left(
\left\{
p_r
\mid
\mathbf{\user}_{r}\in\mathbf{\users}_{j}
\right\}
\right).
\]

The function \(\operatorname{TDProf}\) derives temporal and entity-diversity statistics directly from the profiler records of the users in \(\cluster_j\). The resulting profile \(\Pi^{\mathrm{td}}_j\) summarises temporal activity patterns, activity distributions, contract and token diversity, and entity concentration.

\paragraph{Multi-label category assignment.}
Let \(\mathcal{G}_{\mathrm{tag}}\) denote the set of behavioural and risk-oriented tags used by the profiler. For each cluster \(\cluster_j\), the profiler assigns tags using the user-level evidence of its members and the evidence aggregated for the cluster:
\[
\kappa_j
=
\operatorname{CatProf}
\biggl(
\mathbf{\users}_{j},
\left\{
\bigl(
\pi^{\mathrm{act}}(\mathbf{\user}_{r}),
\pi^{\mathrm{ent}}(\mathbf{\user}_{r}),
\pi^{\mathrm{lab}}(\mathbf{\user}_{r})
\bigr)
\mid
\mathbf{\user}_{r}\in\mathbf{\users}_{j}
\right\},
\Pi^{\mathrm{risk}}_j,
\Delta^{\mathrm{risk}}_j,
\Pi^{\mathrm{tx}}_j,
D^{\mathrm{mot}}_j,
D^{\mathrm{seq}}_j,
\omega_j,
\Pi^{\mathrm{td}}_j
\biggr).
\]

The function \(\operatorname{CatProf}\) compares the available user-level and cluster-level evidence with the behavioural and risk-oriented indicators associated with each candidate tag \(g\in\mathcal{G}_{\mathrm{tag}}\). A tag is assigned when the available evidence provides sufficient support for the corresponding interpretation. The resulting assignment records the strength of the evidence and the behavioural or risk-oriented observations that justify the tag.

The output \(\kappa_j\) summarises the behavioural and risk-oriented tags supported for \(\cluster_j\), together with the strength, supporting evidence, and justification for each assignment:
\[
\kappa_j
=
\left\{
\bigl(
g,
\sigma_j(g),
E_j(g),
J_j(g)
\bigr)
\ \middle|\
g\in\mathcal{G}_{\mathrm{tag}}
\text{ is supported by the available evidence}
\right\}.
\]
For each assigned tag, \(\sigma_j(g)\in[0,1]\) records the strength of the supporting evidence, \(E_j(g)\) identifies the observations supporting the assignment, and \(J_j(g)\) provides a concise justification. The origin of the evidence is retained for traceability. Because \(\kappa_j\) is multi-label, a cluster may receive several behavioural or risk-oriented tags.

The final profile of cluster \(\cluster_j\) is
\[
F_j
=
\operatorname{RenderProf}
\bigl(
\mathbf{\users}_{j},
\kappa_j,
\omega_j,
D^{\mathrm{mot}}_j,
D^{\mathrm{seq}}_j,
\Pi^{\mathrm{td}}_j
\bigr).
\]

The function \(\operatorname{RenderProf}\) organises the cluster membership, assigned tags, semantic reading, defining motifs and sequences, and temporal and diversity profiles into an interpretable cluster report \(F_j\). Behavioural and risk-oriented supporting evidence is carried through \(\kappa_j\), including its evidence, assignment reasons, and provenance.

For each tag \(g\), let
\[
\mathcal{C}_g
=
\left\{
\cluster_j
\mid
g
\text{ is assigned to }
\cluster_j
\right\}
\]
denote the clusters carrying that tag. The tag-level summary is
\[
\Psi(g)
=
\operatorname{TagProf}
\left(
\left\{
F_j
\mid
\cluster_j\in\mathcal{C}_g
\right\}
\right).
\]

The function \(\operatorname{TagProf}\) collects the final profiles of all clusters carrying tag \(g\) and aggregates their recurring motifs, sequences, temporal patterns, entity patterns, and supporting risk evidence into the tag-level summary \(\Psi(g)\). Because a cluster can receive multiple tags, the same cluster profile may contribute to several tag-level summaries.

For risk-oriented tags, \(\sigma_j(g)\) reflects how strongly the available evidence supports a suspicious interpretation, it does not represent the probability that users in the cluster are malicious. Evidence limited to interactions with risky entities or exploit-related transactions is treated as contextual exposure rather than sufficient grounds to characterise the entire cluster as malicious.
\section{Experiments}
\label{sec:experiments}
This section addresses RQ1 and RQ2 by selecting the most suitable framework setting and comparing it with baseline and alternative methods. It compares sentence embeddings, sequence representations, clustering configurations, runtime cost, and pre-labelled address evidence.

\subsection{Runtime Information and Experimental Setup}
Table~\ref{tab:runtime-profile} reports representative runtimes for the main stages of the framework. The dominant costs come from API-dependent preprocessing, especially event decoding and address separation, followed by sentence embedding. In contrast, sentence construction, sequence-model training, sequence embedding generation, and clustering are comparatively lightweight once address artefacts and sentence embeddings have been produced. This supports the practicality of the framework for million-user-scale behavioural analysis, particularly when exsiting data can be reused across runs.

\paragraph{Experimental Setup}
Implementation was developed in Python with PyTorch. Experiments were run on a workstation with a 13th Gen Intel Core i9-13900 CPU, 128GB RAM, and an NVIDIA GeForce RTX 4090 GPU with 24GB memory.

\begin{table}[!t]
\centering
\caption{Typical runtimes for core components of the framework.}
\label{tab:runtime-profile}
\small
\setlength{\tabcolsep}{4pt}
\renewcommand{\arraystretch}{1.08}
\begin{tabular}{@{}
>{\raggedright\arraybackslash}p{0.64\columnwidth}
>{\raggedright\arraybackslash}p{0.28\columnwidth}
@{}}
\hline
\textbf{Stage} & \textbf{Runtime} \\
\hline
Event decoding & 9 hours \\
Address separation & 10 hours \\
ERC-20 enrichment & few minutes \\
ERC-721 enrichment & few minutes \\
Readable ID assignment & 1 minute \\
Build intermediate sentences & 20 minutes \\
Group intermediate sentences & 30 mininutes \\
Build final sentences & 1 hour \\
Group final sentences & few minutes \\
Sentence embedding & 11 hours, 2 models \\
Sequence-model training & \(\sim\)1 minute/model \\
Sequence embedding generation & \(\sim\)1 minute/model \\
Clustering & few minutes \\
\hline
Overall core pipeline & \(\sim\)32 hours \\
\hline
\end{tabular}
\end{table}

\subsection{RQ1: Which Framework Setting is the Most Suitable for Behavioural Analysis?}
This sub-section evaluates the two-step embedding design space on one month of data, covering blocks 16250000 to 16499999. It compares sentence embedding and preprocessing choices, sequence aggregation methods, clustering algorithms and resolutions. The strongest candidates are then assessed through behavioural semantics analysis, since a suitable configuration should recover stable clusters while preserving meaningful behavioural patterns.

\begin{table*}[!t]
\centering
\caption{Sentence embedding model selection and slice-based sanity check. 
\emph{Lower global similarity mean and higher global similarity standard deviation indicate a less-collapsed embedding space.}}
\label{tab:sent-embed-selection}
\small
\setlength{\tabcolsep}{3pt}
\renewcommand{\arraystretch}{1.05}

\begin{tabular*}{\textwidth}{@{\extracolsep{\fill}}llllrrrrrrr@{}}
\hline
\multicolumn{11}{@{}l}{\textbf{(a) Best configuration per sentence embedding model}} \\
\hline
\textbf{Model} &
\textbf{Setting} &
\textbf{Prompt} &
\textbf{Norm} &
\textbf{Score} &
\(\boldsymbol{\mathrm{r@10}}\) &
\(\boldsymbol{\mathrm{AUC}_{act}}\) &
\(\boldsymbol{\mathrm{AUC}_{1}}\) &
\textbf{Margin} &
\textbf{Global \(\boldsymbol{\mu\pm\sigma}\)} &
\textbf{Sent/s} \\
\hline
MiniLM-L6-v2 & \texttt{num\_bucket} & \texttt{none}      & \texttt{post\_norm}   & \num{3.354} & \num{0.724} & \num{0.712} & \num{0.982} & \num{0.445} & \(\num{0.575}\pm\num{0.201}\) & \num{2786} \\
BGE-base     & \texttt{num\_bucket} & \texttt{retrieval} & \texttt{encode\_norm} & \num{3.265} & \num{0.708} & \num{0.720} & \num{0.991} & \num{0.290} & \(\num{0.708}\pm\num{0.123}\) & \num{726}  \\
BGE-small    & \texttt{num\_bucket} & \texttt{retrieval} & \texttt{post\_norm}   & \num{3.241} & \num{0.713} & \num{0.745} & \num{0.989} & \num{0.224} & \(\num{0.796}\pm\num{0.104}\) & \num{1278} \\
GTE-small    & \texttt{num\_bucket} & \texttt{none}      & \texttt{encode\_norm} & \num{3.170} & \num{0.688} & \num{0.756} & \num{0.992} & \num{0.106} & \(\num{0.903}\pm\num{0.046}\) & \num{1272} \\
E5-small     & \texttt{num\_bucket} & \texttt{none}      & \texttt{raw\_dot}     & \num{3.117} & \num{0.689} & \num{0.728} & \num{0.979} & \num{0.089} & \(\num{0.910}\pm\num{0.042}\) & \num{1277} \\
\hline
\end{tabular*}

\vspace{1.5mm}

\begin{tabular*}{\textwidth}{@{\extracolsep{\fill}}lrrrrrrrrrr@{}}
\hline
\multicolumn{11}{@{}l}{\textbf{(b) Slice-based sanity check using \texttt{post\_norm}}} \\
\hline
& \multicolumn{5}{c}{\textbf{Slice A}} & \multicolumn{5}{c}{\textbf{Slice B}} \\
\cline{2-6}\cline{7-11}
\textbf{Model} &
\textbf{Global \(\boldsymbol{\mu\pm\sigma}\)} &
\textbf{Margin} &
\(\boldsymbol{\mathrm{r@10}}\) &
\(\boldsymbol{\mathrm{AUC}_{act}}\) &
\textbf{Sent/s} &
\textbf{Global \(\boldsymbol{\mu\pm\sigma}\)} &
\textbf{Margin} &
\(\boldsymbol{\mathrm{r@10}}\) &
\(\boldsymbol{\mathrm{AUC}_{act}}\) &
\textbf{Sent/s} \\
\hline
MiniLM-L6-v2 & \(\num{0.570}\pm\num{0.197}\) & \num{0.452} & \num{0.723} & \num{0.683} & \num{1724} & \(\num{0.585}\pm\num{0.194}\) & \num{0.437} & \num{0.708} & \num{0.706} & \num{1773} \\
BGE-base     & \(\num{0.704}\pm\num{0.119}\) & \num{0.291} & \num{0.703} & \num{0.721} & \num{633}  & \(\num{0.712}\pm\num{0.118}\) & \num{0.284} & \num{0.691} & \num{0.739} & \num{632}  \\
BGE-small    & \(\num{0.794}\pm\num{0.102}\) & \num{0.224} & \num{0.715} & \num{0.730} & \num{1008} & \(\num{0.801}\pm\num{0.100}\) & \num{0.218} & \num{0.701} & \num{0.759} & \num{1001} \\
\hline
\end{tabular*}
\end{table*}

\subsubsection{Analysis of Embedding Choices at Sentence Level}
\label{sec:sent-embed-selection}
Blockchain transaction sentences are highly templated, so embedding collapse is a key risk: distinct behaviours may be mapped into a narrow similarity band. We therefore evaluated sentence embedding models using weak-label diagnostics that measure both semantic discrimination and geometric spread, including local ranking quality \(r@10\), alias-level separation \(\mathrm{AUC}_{1}\), activity-level separation \(\mathrm{AUC}_{\mathrm{act}}\), separation margin, and global similarity mean and standard deviation. These diagnostics assess representation quality rather than malicious/benign classification. Positive pairs are constructed from shared action aliases, token signatures, or inferred activity categories such as swap, borrow, transfer, and approval, excluding broad ``other'' cases from activity-level AUC computation.

We compared MiniLM-L6-v2~\citep{wang2020minilm,reimers2019sentence}, BGE-base and BGE-small~\citep{xiao2023bge}, GTE-small~\citep{li2023gte}, and E5-small~\citep{wang2022text} across six preprocessing settings, six prompt variants, and three normalisation modes, using up to \num{50000} users, \num{12000} sentences, at most six sentences per user, batch size 1024, and \texttt{bf16} precision. As shown in Table~\ref{tab:sent-embed-selection}, MiniLM-L6-v2 with \texttt{num\_bucket} preprocessing, no prompt, and \texttt{post\_norm} achieved the best overall profile, with the lowest global similarity mean, highest global similarity spread, and largest separation margin, while maintaining strong \(r@10\) and AUC values. The slice-based sanity check in Table~\ref{tab:sent-embed-selection}(b) supports this ranking: MiniLM-L6-v2 preserved its expanded-geometry pattern across slices, while BGE-base and BGE-small were stable but more compressed. We therefore use MiniLM-L6-v2 with \texttt{num\_bucket} preprocessing and no prompt as the primary sentence embedding configuration, retaining BGE-small and BGE-base with \texttt{num\_bucket} and the retrieval prompt as robustness alternatives.

\subsubsection{Analysis of Embedding Choices at Sequence Level}
\label{sec:seq-embed-selection}
After selecting sentence encoders, we evaluated how each user's sentence sequence should be aggregated into a single behavioural vector. The selection process used two checks: a multi-\(k\), multi-seed clustering audit to measure cluster quality, stability, cross-\(k\) smoothness, balance, and runtime; and a leakage audit to test whether clusters were mainly driven by trivial signals such as sequence length or activity volume.

\begin{table*}[pos=!t,width=\textwidth]
\centering
\caption{Sequence aggregation model selection and sanity-audit results.
\emph{Step~1 scores are reported as mean$\pm$std across
\(k\in\{\num{15},\num{20},\num{25},\num{50},\num{100}\}\)
and \num{5} seeds. Lower median row time indicates higher throughput.
Lower \(\eta^2\) gap indicates less sequence-length leakage.}}
\label{tab:seq-aggregation-selection}

\small
\setlength{\tabcolsep}{3pt}
\renewcommand{\arraystretch}{1.05}

\makebox[\textwidth][c]{%
\begin{tabular}{@{}
Y{0.10\textwidth}
Y{0.16\textwidth}
Y{0.09\textwidth}
Y{0.17\textwidth}
Y{0.12\textwidth}
Y{0.05\textwidth}
@{}}
\hline
\multicolumn{6}{@{}l}{%
\textbf{(a) Top Step~1 configurations ranked by aggregated clustering score}} \\
\hline
\textbf{Seq. model} &
\textbf{Sent. model} &
\textbf{Norm} &
\textbf{Step-1 score} &
\textbf{Row time (s)} &
\textbf{Tier} \\
\hline
CNN    & BGE-base     & \texttt{raw} & \(\num{0.962}\pm\num{0.035}\) & \num{91.3} & A \\
GRU    & BGE-base     & \texttt{raw} & \(\num{0.853}\pm\num{0.015}\) & \num{94.4} & A \\
GRU    & BGE-base     & L2           & \(\num{0.838}\pm\num{0.007}\) & \num{97.7} & B \\
CNN    & MiniLM-L6-v2 & L2           & \(\num{0.818}\pm\num{0.037}\) & \num{42.6} & B \\
Mamba2 & BGE-small    & L2           & \(\num{0.803}\pm\num{0.028}\) & \num{43.7} & B \\
CNN    & MiniLM-L6-v2 & \texttt{raw} & \(\num{0.786}\pm\num{0.036}\) & \num{42.7} & C \\
Mean   & BGE-base     & \texttt{raw} & \(\num{0.782}\pm\num{0.024}\) & \num{91.8} & C \\
GRU    & MiniLM-L6-v2 & \texttt{raw} & \(\num{0.731}\pm\num{0.037}\) & \num{42.8} & C \\
Mean   & BGE-base     & L2           & \(\num{0.740}\pm\num{0.012}\) & \num{92.1} & C \\
GRU    & BGE-small    & \texttt{raw} & \(\num{0.712}\pm\num{0.071}\) & \num{42.3} & C \\
\hline
\end{tabular}%
}

\vspace{1.5mm}

\makebox[\textwidth][c]{%
\begin{tabular}{@{}
Y{0.1\textwidth}
Y{0.12\textwidth}
Y{0.08\textwidth}
Y{0.12\textwidth}
Y{0.035\textwidth}
Y{0.12\textwidth}
Y{0.035\textwidth}
Y{0.075\textwidth}
Y{0.075\textwidth}
Y{0.07\textwidth}
Y{0.04\textwidth}
@{}}
\hline
\multicolumn{11}{@{}l}{%
\textbf{(b) Combined Step~1 robustness and Step~2 sanity summary}} \\
\hline
\textbf{Seq. model} &
\textbf{Sent. model} &
\textbf{Norm} &
\textbf{Step-1} &
\textbf{T1} &
\textbf{Sanity} &
\textbf{T2} &
\(\boldsymbol{\eta^2_{\mathrm{len}}}\) &
\(\boldsymbol{\eta^2}\) \textbf{gap} &
\textbf{Overall} &
\textbf{Tier} \\
\hline
Mamba2 & BGE-small    & L2           & \(\num{0.803}\pm\num{0.028}\) & B & \(\num{0.530}\pm\num{0.194}\) & C & \num{0.344} & \num{0.344} & \num{0.817} & A \\
CNN    & MiniLM-L6-v2 & L2           & \(\num{0.818}\pm\num{0.037}\) & B & \(\num{0.316}\pm\num{0.048}\) & C & \num{0.603} & \num{0.603} & \num{0.793} & A \\
GRU    & MiniLM-L6-v2 & \texttt{raw} & \(\num{0.731}\pm\num{0.037}\) & C & \(\num{0.527}\pm\num{0.164}\) & C & \num{0.309} & \num{0.309} & \num{0.751} & B \\
Mean   & MiniLM-L6-v2 & L2           & \(\num{0.749}\pm\num{0.010}\) & D & \(\num{0.388}\pm\num{0.027}\) & C & \num{0.336} & \num{0.336} & \num{0.614} & C \\
GRU    & BGE-small    & \texttt{raw} & \(\num{0.712}\pm\num{0.071}\) & C & \(\num{0.340}\pm\num{0.171}\) & D & \num{0.300} & \num{0.300} & \num{0.610} & C \\
GRU    & MiniLM-L6-v2 & L2           & \(\num{0.680}\pm\num{0.003}\) & D & \(\num{0.492}\pm\num{0.076}\) & B & \num{0.277} & \num{0.277} & \num{0.596} & C \\
GRU    & BGE-small    & L2           & \(\num{0.621}\pm\num{0.052}\) & E & \(\num{0.460}\pm\num{0.074}\) & B & \num{0.308} & \num{0.308} & \num{0.483} & D \\
Mamba2 & MiniLM-L6-v2 & L2           & \(\num{0.680}\pm\num{0.020}\) & D & \(\num{0.331}\pm\num{0.102}\) & D & \num{0.328} & \num{0.328} & \num{0.365} & D \\
\hline
\end{tabular}%
}

\end{table*}

\paragraph{Step 1: Multi-\(k\) Clustering Audit}
We evaluated mean pooling, GRU~\citep{cho2014gru}, TX~\citep{vaswani2017attention}, CNN~\citep{kim2014cnn}, and Mamba2~\citep{dao2024mamba2} over MiniLM-L6-v2, BGE-small, and BGE-base sentence embeddings, using both \texttt{raw} and L2-normalised user vectors. MiniBatch \textit{K}-means was run with \(k\in\{15,20,25,50,100\}\) and five random seeds, measuring clustering quality, seed stability, cross-\(k\) alignment, centroid drift, cluster balance, and speed. Table~\ref{tab:seq-aggregation-selection}(a) shows that CNN and GRU with BGE-base achieved the strongest Step~1 scores, with CNN \(\times\) BGE-base \(\times\) \texttt{raw} ranked first. However, BGE-base was substantially slower than MiniLM-L6-v2 and BGE-small. Among the efficient configurations, CNN \(\times\) MiniLM-L6-v2 \(\times\) L2, Mamba2 \(\times\) BGE-small \(\times\) L2, and GRU variants with MiniLM-L6-v2 or BGE-small remained competitive.

\paragraph{Step 2: Sanity Audit}
Since strong clustering scores can reflect trivial activity signals, we tested whether clusters were mainly driven by sequence length or activity volume. The sanity audit focused on the more cost-efficient sentence models, MiniLM-L6-v2 and BGE-small, and measured length-to-cluster association using \(\eta^2\), mutual information, within-bin dominance, effective cluster count, and permuted-label baselines. Table~\ref{tab:seq-aggregation-selection}(b) shows that CNN retained strong Step~1 performance but also had substantial length leakage, especially CNN \(\times\) MiniLM-L6-v2 \(\times\) L2 with an \(\eta^2\) gap of \(\num{0.603}\). GRU variants had the lowest leakage, with \(\eta^2\) gaps around \(\num{0.277}\) to \(\num{0.309}\), while Mean and Mamba2 were intermediate. We therefore carried forward GRU configurations with MiniLM-L6-v2 and BGE-small under both \texttt{raw} and L2 normalisation, kept Mamba2 \(\times\) MiniLM-L6-v2 \(\times\) L2 as a complementary model, and retained Mean \(\times\) MiniLM-L6-v2 \(\times\) L2 as a simple baseline. CNN was excluded from the behavioural analysis because of length leakage, and BGE-base was not prioritised due to higher runtime and lack of sanity-audit coverage.

\begin{table*}[!t]
\centering
\caption{Training-objective and sequence-length sweeps for selected learned sequence models. \emph{Panel (a) reports the top two objectives for each selected learned setting with \(\texttt{max\_len}=\num{64}\), with ranges over \(k\in\{\num{15},\num{20},\num{25},\num{50},\num{100}\}\). Panel (b) reports the best length and cropping configuration for each selected target. Lower \(\mathrm{act}\_\eta^2\) indicates less activity-volume confounding.}}
\label{tab:seq-objective-length}
\small
\setlength{\tabcolsep}{2pt}
\renewcommand{\arraystretch}{1.05}

\begin{tabular}{@{}
>{\raggedright\arraybackslash}p{0.15\textwidth}
>{\raggedright\arraybackslash}p{0.1\textwidth}
>{\raggedright\arraybackslash}p{0.07\textwidth}
>{\raggedright\arraybackslash}p{0.10\textwidth}
>{\raggedright\arraybackslash}p{0.14\textwidth}
>{\raggedright\arraybackslash}p{0.135\textwidth}
>{\raggedright\arraybackslash}p{0.055\textwidth}
>{\raggedright\arraybackslash}p{0.075\textwidth}
@{}}
\hline
\multicolumn{8}{@{}l}{\textbf{(a) Training objective sweep}} \\
\hline
\textbf{Sentence model} &
\textbf{Seq. model} &
\textbf{Norm} &
\textbf{Objective} &
\textbf{SC range} &
\textbf{stabARI range} &
\textbf{DBI} &
\textbf{maxFrac} \\
\hline
MiniLM-L6-v2 & GRU    & L2           & \texttt{masked}   & \num{0.454}--\num{0.480} & \num{0.771}--\num{0.922} & \num{1.354} & \num{0.298} \\
MiniLM-L6-v2 & GRU    & L2           & \texttt{nextstep} & \num{0.413}--\num{0.452} & \num{0.849}--\num{0.962} & \num{1.507} & \num{0.331} \\
MiniLM-L6-v2 & GRU    & \texttt{raw} & \texttt{nextstep} & \num{0.400}--\num{0.429} & \num{0.832}--\num{0.959} & \num{1.568} & \num{0.322} \\
MiniLM-L6-v2 & GRU    & \texttt{raw} & \texttt{masked}   & \num{0.189}--\num{0.458} & \num{0.841}--\num{0.927} & \num{1.403} & \num{0.315} \\
BGE-small    & GRU    & L2           & \texttt{masked}   & \num{0.462}--\num{0.494} & \num{0.854}--\num{0.969} & \num{1.349} & \num{0.307} \\
BGE-small    & GRU    & L2           & \texttt{nextstep} & \num{0.436}--\num{0.463} & \num{0.864}--\num{0.964} & \num{1.455} & \num{0.312} \\
BGE-small    & GRU    & \texttt{raw} & \texttt{masked}   & \num{0.445}--\num{0.518} & \num{0.822}--\num{0.972} & \num{1.305} & \num{0.302} \\
BGE-small    & GRU    & \texttt{raw} & \texttt{nextstep} & \num{0.443}--\num{0.484} & \num{0.838}--\num{0.933} & \num{1.418} & \num{0.288} \\
MiniLM-L6-v2 & Mamba2 & L2           & \texttt{nextstep} & \num{0.409}--\num{0.429} & \num{0.806}--\num{0.961} & \num{1.665} & \num{0.319} \\
MiniLM-L6-v2 & Mamba2 & L2           & \texttt{masked}   & \num{0.406}--\num{0.434} & \num{0.866}--\num{0.957} & \num{1.914} & \num{0.324} \\
\hline
\end{tabular}

\vspace{1.5mm}

\begin{tabular}{@{}
>{\raggedright\arraybackslash}p{0.34\textwidth}
>{\centering\arraybackslash}p{0.1\textwidth}
>{\raggedright\arraybackslash}p{0.30\textwidth}
>{\centering\arraybackslash}p{0.105\textwidth}
>{\centering\arraybackslash}p{0.055\textwidth}
>{\centering\arraybackslash}p{0.07\textwidth}
@{}}
\hline
\multicolumn{6}{@{}l}{\textbf{(b) Length and cropping sweep}} \\
\hline
\textbf{Target} &
\textbf{Max len.} &
\textbf{Sampling setting} &
\textbf{Score} &
\textbf{ARI} &
\(\boldsymbol{\mathrm{act}\_\eta^2}\) \\
\hline

GRU \texttt{masked} + MiniLM-L6-v2 (L2) &
\num{32} &
\texttt{prefix}, \texttt{sqrt\_len} &
\(\num{0.301}\pm\num{0.026}\) &
\num{0.921} &
\num{0.00085} \\

GRU \texttt{nextstep} + MiniLM-L6-v2 (\texttt{raw}) &
\num{128} &
\texttt{prefix}, \texttt{uniform} &
\(\num{0.241}\pm\num{0.015}\) &
\num{0.890} &
\num{0.00061} \\

GRU \texttt{masked} + BGE-small (L2) &
\num{128} &
\texttt{stratified\_window}, \texttt{sqrt\_len} &
\(\num{0.319}\pm\num{0.020}\) &
\num{0.917} &
\num{0.00098} \\

GRU \texttt{masked} + BGE-small (\texttt{raw}) &
\num{32} &
\texttt{stratified\_window}, \texttt{sqrt\_len} &
\(\num{0.315}\pm\num{0.031}\) &
\num{0.892} &
\num{0.00103} \\

Mamba2 \texttt{nextstep} + MiniLM-L6-v2 (L2) &
\num{32} &
\texttt{prefix}, \texttt{uniform} &
\(\num{0.218}\pm\num{0.018}\) &
\num{0.879} &
\num{0.00069} \\
\hline
\end{tabular}
\end{table*}

\paragraph{Learned Sequence Models: Training Objective Sweep and Length Optimisation}
After selecting the main sequence aggregators, we trained compact GRU and Mamba2 models to capture behavioural dynamics from sentence embeddings. The sweep covered GRU \(\times\) MiniLM-L6-v2 with \texttt{raw}/L2 normalisation, GRU \(\times\) BGE-small with \texttt{raw}/L2 normalisation, and Mamba2 \(\times\) MiniLM-L6-v2 with L2 normalisation. Training used \(\texttt{max\_len}=\num{64}\), \(\texttt{batch\_users\_train}=\num{256}\), 2000 GPU steps with AMP, and MiniBatch \textit{K}-means evaluation over \(k\in\{\num{15},\num{20},\num{25},\num{50},\num{100}\}\) with five seeds. Table~\ref{tab:seq-objective-length}(a) shows that the best objective was setting-dependent: \texttt{masked} was strongest for most GRU configurations, while GRU \(\times\) MiniLM-L6-v2 with \texttt{raw} normalisation and Mamba2 \(\times\) MiniLM-L6-v2 with L2 favoured \texttt{nextstep}.

We also swept sequence length, cropping, and sampling settings to justify bounded histories. In the first-month data, \num{6045497} users had median sequence length \(\num{2}\), \(p90=\num{9}\), and \(p99=\num{76}\), but a long tail up to \(\num{7.46e5}\) events. Table~\ref{tab:seq-objective-length}(b) shows that the best setting was target-dependent: GRU variants preferred either \(\texttt{max\_len}=\num{32}\) or \(\num{128}\), with \texttt{prefix} or \texttt{stratified\_window} cropping depending on the sentence model and normalisation, while Mamba2 performed best with \(\texttt{max\_len}=\num{32}\), \texttt{prefix} cropping, and \texttt{uniform} sampling. These capped settings cover most users while limiting compute and reducing domination by extreme-length accounts.

\paragraph{Final Configurations Based on Structural Evidence}
Based on the objective and length sweeps, we proceeded with six configurations:
(i) a non-learned baseline using mean pooling over MiniLM-L6-v2 embeddings with L2 normalisation;
(ii) GRU \(\times\) MiniLM-L6-v2 \(\times\) \texttt{masked} \(\times\) L2 with \texttt{max\_len}=\num{32} (\texttt{prefix}, \texttt{sqrt\_len});
(iii) GRU \(\times\) MiniLM-L6-v2 \(\times\) \texttt{nextstep} \(\times\) \texttt{raw} with \texttt{max\_len}=\num{128} (\texttt{prefix}, \texttt{uniform});
(iv) GRU \(\times\) BGE-small \(\times\) \texttt{masked} \(\times\) L2 with \texttt{max\_len}=\num{128} (\texttt{stratified\_window}, \texttt{sqrt\_len});
(v) GRU \(\times\) BGE-small \(\times\) \texttt{masked} \(\times\) \texttt{raw} with \texttt{max\_len}=\num{32} (\texttt{stratified\_window}, \texttt{sqrt\_len}); and
(vi) Mamba2 \(\times\) MiniLM-L6-v2 \(\times\) \texttt{nextstep} \(\times\) L2 with \texttt{max\_len}=\num{32} (\texttt{prefix}, \texttt{uniform}).

\begin{table*}[!t]
\centering
\caption{\(k\)-sweep peaks, stability, and elbow-search caps. \emph{Metrics are reported as mean$\pm$std over \(R=\num{5}\) runs on \(X_{\mathrm{eval}}\). \(k^{\star}\) denotes the best \(k\) from the sweep, and \(k_{\max}\) denotes the maximum \(k\) retained for the subsequent elbow search.}}
\label{tab:k_sweep_summary}
\small
\setlength{\tabcolsep}{4pt}
\renewcommand{\arraystretch}{1.08}
\begin{tabular}{@{}
>{\raggedright\arraybackslash}p{0.36\textwidth}
c
c
>{\centering\arraybackslash}p{0.15\textwidth}
>{\centering\arraybackslash}p{0.15\textwidth}
>{\centering\arraybackslash}p{0.13\textwidth}
@{}}
\hline
\textbf{Configuration} &
\(\boldsymbol{k^{\star}}\) &
\(\boldsymbol{k_{\max}}\) &
\textbf{SC} &
\textbf{DBI} &
\textbf{NMI / AMI} \\
\hline

Mean \(\times\) MiniLM-L6-v2 (L2) &
\num{10} &
\num{30} &
\(\num{0.4385}\pm\num{0.0129}\) &
\(\num{1.8210}\pm\num{0.0493}\) &
\(\num{0.7932}/\num{0.7931}\) \\

GRU \(\times\) MiniLM-L6-v2 (\texttt{masked}, L2) &
\num{10} &
\num{60} &
\(\num{0.4881}\pm\num{0.0131}\) &
\(\num{1.2516}\pm\num{0.0636}\) &
\(\num{0.8456}/\num{0.8456}\) \\

GRU \(\times\) MiniLM-L6-v2 (\texttt{nextstep}, \texttt{raw}) &
\num{60} &
\num{80} &
\(\num{0.4372}\pm\num{0.0045}\) &
\(\num{1.5896}\pm\num{0.0466}\) &
\(\num{0.8230}/\num{0.8225}\) \\

GRU \(\times\) BGE-small (\texttt{masked}, L2) &
\num{20} &
\num{60} &
\(\num{0.4813}\pm\num{0.0120}\) &
\(\num{1.3537}\pm\num{0.0370}\) &
\(\num{0.8321}/\num{0.8321}\) \\

GRU \(\times\) BGE-small (\texttt{masked}, \texttt{raw}) &
\num{100} &
\num{140} &
\(\num{0.5104}\pm\num{0.0098}\) &
\(\num{1.3888}\pm\num{0.0172}\) &
\(\num{0.8535}/\num{0.8525}\) \\

Mamba2 \(\times\) MiniLM-L6-v2 (\texttt{nextstep}, L2) &
\num{10} &
\num{60} &
\(\num{0.4387}\pm\num{0.0340}\) &
\(\num{1.4681}\pm\num{0.0522}\) &
\(\num{0.7952}/\num{0.7952}\) \\
\hline
\end{tabular}
\end{table*}

\subsubsection{Analysis of Clustering Configurations}
To select the clustering configuration for downstream behavioural analysis, we first ran a \(k\)-sweep consistency study for each embedding configuration and then compared several clustering algorithms at the selected resolutions. The comparison included full \textit{K}-means~\citep{macqueen1967some}, MiniBatch \textit{K}-means~\citep{10.1145/1772690.1772862}, FAISS \textit{K}-means~\citep{faiss2025}, Leiden community detection~\citep{traag2019louvain}, and hybrid \textit{K}-means+Leiden refinement. Because many users had only one transaction, the strongest \textit{K}-means settings were also re-tested after excluding length-\(1\) users.

\paragraph{\(k\)-sweep Consistency Study}
Per configuration, we swept \(k \in \{\num{10},\num{20},\num{30},\num{40},\num{60},\num{80},\num{100},\num{140},\num{200},\num{300},\num{450},\num{650}\}\). Each setting was evaluated over \(R=5\) runs using an evaluation set \(X_{\mathrm{eval}}\) of \num{200000} users and a fresh training set \(X_{\mathrm{train}}\) of the same size per run. MiniBatch \textit{K}-means was fitted on \(X_{\mathrm{train}}\), labels were predicted on \(X_{\mathrm{eval}}\), and clustering quality and stability metrics were recorded. Run-to-run consistency was measured using pairwise NMI and AMI.

Table~\ref{tab:k_sweep_summary} reports the best \(k\) for each configuration, selected by mean SC, together with DBI, NMI/AMI stability, and the \(k_{\max}\) cap used for the subsequent elbow search. The results indicate that the preferred clustering resolution is configuration-dependent rather than governed by a single universal \(k\).

\begin{table*}[!t]
\centering
\caption{Clustering method selection and selected configuration summary. \emph{Panel (a) reports the best retained clustering method for each candidate embedding configuration. Panel (b) reports the selected configurations before and after excluding length-\(\num{1}\) users. Higher SC and CHI are better; lower DBI, maximum cluster fraction, Gini, length leakage, and number of clusters below \(\num{1}\%\) are preferred. ``All'' denotes all users, while \(\geq\num{2}\) excludes users with only one linked transaction. The column \(<1\%\) gives the number of clusters containing fewer than \(\num{1}\%\) of users.}}
\label{tab:clustering-selection-summary}
\small
\setlength{\tabcolsep}{2pt}
\renewcommand{\arraystretch}{1.06}

\begin{tabular}{@{}
>{\raggedright\arraybackslash}p{0.10\textwidth}
>{\raggedright\arraybackslash}p{0.28\textwidth}
>{\raggedright\arraybackslash}p{0.22\textwidth}
>{\centering\arraybackslash}p{0.09\textwidth}
>{\centering\arraybackslash}p{0.09\textwidth}
>{\centering\arraybackslash}p{0.11\textwidth}
@{}}
\hline
\multicolumn{6}{@{}l}{\textbf{(a) Best clustering method per candidate embedding configuration}} \\
\hline
\textbf{Embedding} &
\textbf{Configuration} &
\textbf{Selected method} &
\textbf{SC} &
\textbf{DBI} &
\textbf{CHI} \\
\hline

Mean &
MiniLM-L6-v2 (L2) &
MiniBatch \textit{K}-means, refined &
\num{0.437} & \num{1.87} & \num{7747} \\

GRU &
MiniLM-L6-v2, \texttt{masked}, L2 &
Full \textit{K}-means, cosine &
\num{0.468} & \num{1.411} & \num{21260} \\

GRU &
MiniLM-L6-v2, \texttt{nextstep}, \texttt{raw} &
MiniBatch \textit{K}-means, refined &
\num{0.371} & \num{1.651} & \num{7206} \\

GRU &
BGE-small, \texttt{masked}, L2 &
Full \textit{K}-means, cosine &
\num{0.465} & \num{1.309} & \num{26559} \\

GRU &
BGE-small, \texttt{masked}, \texttt{raw} &
Full \textit{K}-means, cosine &
\num{0.427} & \num{1.373} & \num{11689} \\

Mamba2 &
MiniLM-L6-v2, \texttt{nextstep}, L2 &
Full \textit{K}-means, cosine &
\num{0.414} & \num{1.65} & \num{11043} \\
\hline
\end{tabular}

\vspace{1.5mm}

{
\setlength{\tabcolsep}{1.4pt}
\begin{tabular}{@{}
>{\raggedright\arraybackslash}p{0.065\textwidth}
>{\raggedright\arraybackslash}p{0.245\textwidth}
>{\centering\arraybackslash}p{0.055\textwidth}
>{\centering\arraybackslash}p{0.035\textwidth}
>{\centering\arraybackslash}p{0.055\textwidth}
>{\centering\arraybackslash}p{0.055\textwidth}
>{\centering\arraybackslash}p{0.070\textwidth}
>{\centering\arraybackslash}p{0.055\textwidth}
>{\centering\arraybackslash}p{0.060\textwidth}
>{\centering\arraybackslash}p{0.055\textwidth}
>{\centering\arraybackslash}p{0.070\textwidth}
>{\centering\arraybackslash}p{0.050\textwidth}
@{}}
\hline
\multicolumn{12}{@{}l}{\textbf{(b) Selected clustering configurations}} \\
\hline
\textbf{Model} &
\textbf{Setting} &
\textbf{Users} &
\(\boldsymbol{k}\) &
\textbf{SC} &
\textbf{DBI} &
\textbf{CHI} &
\textbf{Max} &
\textbf{Min} &
\textbf{Gini} &
\(\boldsymbol{\eta^2_{\mathrm{len}}}\) &
\(\boldsymbol{<\!1\%}\) \\
\hline

Mean &
MiniLM-L6-v2 (L2) &
All &
\num{20} &
\num{0.437} &
\num{1.872} &
\num{7747.3} &
\num{0.362} &
\num{0.0098} &
\num{0.532} &
\num{0.315} &
\num{2} \\

Mean &
MiniLM-L6-v2 (L2) &
\(\geq\num{2}\) &
\num{22} &
\num{0.329} &
\num{2.132} &
\num{4382.1} &
\num{0.165} &
\num{0.0202} &
\num{0.324} &
\num{0.196} &
\num{0} \\

GRU &
MiniLM-L6-v2, \texttt{masked}, L2 &
All &
\num{26} &
\num{0.468} &
\num{1.411} &
\num{21260.3} &
\num{0.355} &
\num{0.0065} &
\num{0.513} &
\num{0.313} &
\num{2} \\

GRU &
MiniLM-L6-v2, \texttt{masked}, L2 &
\(\geq\num{2}\) &
\num{19} &
\num{0.438} &
\num{1.481} &
\num{12405.3} &
\num{0.176} &
\num{0.0188} &
\num{0.331} &
\num{0.175} &
\num{0} \\

GRU &
BGE-small, \texttt{masked}, L2 &
All &
\num{25} &
\num{0.465} &
\num{1.309} &
\num{26558.9} &
\num{0.358} &
\num{0.0040} &
\num{0.576} &
\num{0.316} &
\num{4} \\

GRU &
BGE-small, \texttt{masked}, L2 &
\(\geq\num{2}\) &
\num{26} &
\num{0.420} &
\num{1.524} &
\num{11579.0} &
\num{0.183} &
\num{0.0109} &
\num{0.395} &
\num{0.200} &
\num{0} \\

Mamba2 &
MiniLM-L6-v2, \texttt{nextstep}, L2 &
All &
\num{29} &
\num{0.414} &
\num{1.648} &
\num{11043.1} &
\num{0.380} &
\num{0.0053} &
\num{0.553} &
\num{0.288} &
\num{5} \\

Mamba2 &
MiniLM-L6-v2, \texttt{nextstep}, L2 &
\(\geq\num{2}\) &
\num{22} &
\num{0.336} &
\num{1.695} &
\num{8544.8} &
\num{0.143} &
\num{0.0137} &
\num{0.315} &
\num{0.171} &
\num{0} \\
\hline
\end{tabular}
}
\end{table*}

\paragraph{Clustering Algorithm Selection}
After fixing a reasonable clustering resolution for each representation using the \(k\)-sweep, we compared practical clustering algorithms at the selected resolutions. 
To ensure comparability, clustering used cosine distance after PCA reduction to 128 dimensions. The PCA transform was trained on \num{400000} sampled users, and clustering metrics were computed on a consistent evaluation sample of \num{200000} users. Each final \(k\) was selected by an inertia-based elbow search capped by the preceding \(k\)-sweep.

Table~\ref{tab:clustering-selection-summary}(a) summarises the best retained clustering method for each candidate embedding configuration. Across the selected embedding spaces, the \textit{K}-means family produced the most reliable geometric structure: Leiden-only clustering fragmented the space, and hybrid \textit{K}-means+Leiden refinement did not improve the trade-off. Full \textit{K}-means was retained for the learned GRU and Mamba2 embeddings, while refined MiniBatch \textit{K}-means was retained for the mean baseline. The strongest geometric results came from \texttt{masked} GRU embeddings: GRU \(\times\) MiniLM-L6-v2 \(\times\) L2 achieved the highest SC, while GRU \(\times\) BGE-small \(\times\) L2 achieved the lowest DBI and highest CHI. The \texttt{raw} GRU variants were weaker and more length-confounded, whereas Mamba2 was retained as a complementary option because it showed lower length leakage.

A focused follow-up excluded users with only one linked transaction, reducing the population from \num{6045497} to \num{3032235}. As shown in Table~\ref{tab:clustering-selection-summary}(b), the \texttt{masked} GRU configurations remained strongest on SC, DBI, and CHI, while the filtered partitions became more balanced and less length-driven, with no clusters below \(\num{1}\%\) of users. The carried-forward configurations use \(k=\num{20}\) for the mean baseline, \(k=\num{26}\) for GRU MiniLM, \(k=\num{25}\) for GRU BGE-small, and \(k=\num{29}\) for Mamba2 in the full-population setting; after filtering, the corresponding resolutions are \(k=\num{22}\), \(k=\num{19}\), \(k=\num{26}\), and \(k=\num{22}\). These selected configurations form the input to the behavioural semantic analysis stage.

\subsubsection{Behavioural Semantics Analysis}
The previous tests identified structurally strong clustering candidates, but the final setting should also preserve behavioural semantics. We therefore compared the selected configurations by measuring how well they concentrate pre-labelled phishing, other malicious, and bot addresses within individual clusters. The same \num{175} labelled addresses were used throughout: the full-population setting clustered \num{6045497} users and recovered \num{117} labelled addresses, while the length \(\geq\num{2}\) setting clustered \num{3032235} users and recovered \num{108} labelled addresses.

Table~\ref{tab:prelabelled_cluster_concentration} compares the strongest observed label concentrations produced by each configuration. These values are used as descriptive selection signals rather than as performance metrics. In the full-population setting, GRU \(\times\) BGE-small \(\times\) \texttt{masked} \(\times\) L2 gave the strongest phishing concentration, while Mamba2 gave the strongest bot concentration. In the length \(\geq\num{2}\) setting, Mamba2 achieved the strongest phishing concentration, with \(\num{15}/\num{32}\) phishing addresses (\(\num{46.9}\%\)) in one cluster, while GRU \(\times\) BGE-small \(\times\) \texttt{masked} \(\times\) L2 gave the strongest bot concentration. The other-malicious sample is small, with only \(\num{9}\) matched addresses, so those results are interpreted cautiously. Combining these concentration results with the structural results in Table~\ref{tab:clustering-selection-summary}(b), we select the length \(\geq\num{2}\) Mamba2 \(\times\) MiniLM-L6-v2 \(\times\) \texttt{nextstep} \(\times\) L2 setting for downstream behavioural analysis. It uses \(\texttt{max\_len}=\num{32}\), \texttt{prefix} cropping, \texttt{uniform} user sampling, PCA-\num{128}, cosine full \textit{K}-means, and \(k=\num{22}\); future runs can reselect \(k\) using the elbow method with \(k_{\max}=\num{30}\).

\begin{table*}[!t]
\centering
\caption{Pre-labelled behavioural concentration across the selected clustering configurations. \emph{``Other mal.'' combines labels such as bug exploit, front run, and oracle manipulation. For phishing, other malicious, and bots, the table reports the largest count observed in any single cluster. ``All'' denotes all users, while \(\geq\num{2}\) excludes users with only one linked transaction.}}
\label{tab:prelabelled_cluster_concentration}
\small
\setlength{\tabcolsep}{3pt}
\renewcommand{\arraystretch}{1.08}

\begin{tabular}{@{}lcccccc@{}}
\hline
\textbf{Configuration} &
\textbf{Users} &
\textbf{Labelled found} &
\textbf{Phishing} &
\textbf{Other mal.} &
\textbf{Bots} &
\textbf{Bot spread} \\
\hline

Mean \(\times\) MiniLM-L6-v2 (L2) &
All &
\(\num{117}/\num{175}\) &
\(\num{10}/\num{38}\) &
\(\num{3}/\num{9}\) &
\(\num{20}/\num{70}\) &
\num{12} \\

Mean \(\times\) MiniLM-L6-v2 (L2) &
\(\geq\num{2}\) &
\(\num{108}/\num{175}\) &
\(\num{10}/\num{32}\) &
\(\num{2}/\num{9}\) &
\(\num{19}/\num{67}\) &
\num{15} \\

GRU \(\times\) MiniLM-L6-v2 \(\times\) \texttt{masked} \(\times\) L2 &
All &
\(\num{117}/\num{175}\) &
\(\num{13}/\num{38}\) &
\(\num{3}/\num{9}\) &
\(\num{21}/\num{70}\) &
\num{18} \\

GRU \(\times\) MiniLM-L6-v2 \(\times\) \texttt{masked} \(\times\) L2 &
\(\geq\num{2}\) &
\(\num{108}/\num{175}\) &
\(\num{14}/\num{32}\) &
\(\num{2}/\num{9}\) &
\(\num{18}/\num{67}\) &
\num{15} \\

GRU \(\times\) BGE-small \(\times\) \texttt{masked} \(\times\) L2 &
All &
\(\num{117}/\num{175}\) &
\(\num{22}/\num{38}\) &
\(\num{2}/\num{9}\) &
\(\num{20}/\num{70}\) &
\num{13} \\

GRU \(\times\) BGE-small \(\times\) \texttt{masked} \(\times\) L2 &
\(\geq\num{2}\) &
\(\num{108}/\num{175}\) &
\(\num{12}/\num{32}\) &
\(\num{2}/\num{9}\) &
\(\num{20}/\num{67}\) &
\num{17} \\

Mamba2 \(\times\) MiniLM-L6-v2 \(\times\) \texttt{nextstep} \(\times\) L2 &
All &
\(\num{117}/\num{175}\) &
\(\num{10}/\num{38}\) &
\(\num{2}/\num{9}\) &
\(\num{22}/\num{70}\) &
\num{21} \\

Mamba2 \(\times\) MiniLM-L6-v2 \(\times\) \texttt{nextstep} \(\times\) L2 &
\(\geq\num{2}\) &
\(\num{108}/\num{175}\) &
\(\num{15}/\num{32}\) &
\(\num{2}/\num{9}\) &
\(\num{18}/\num{67}\) &
\num{17} \\
\hline
\end{tabular}%
\end{table*}

\begin{tcolorbox}[rqbox,title=Answer to RQ1]
The length \(\geq\num{2}\) Mamba2 \(\times\) MiniLM-L6-v2 \(\times\) \texttt{nextstep} \(\times\) L2 setting is selected for behavioural analysis, using \(\texttt{max\_len}=\num{32}\), \texttt{prefix} cropping, \texttt{uniform} sampling, PCA-\num{128}, cosine full \textit{K}-means, and \(k=\num{22}\). Although the \texttt{masked} GRU variants have stronger geometric scores, Mamba2 gives the best filtered phishing concentration and the cleanest partition, with lower cluster imbalance and weaker length leakage.
\end{tcolorbox}

\subsection{RQ2: How Does the Proposed Framework Perform Compared with Baseline and Alternative Methods?}
We compare the selected framework configuration with two baselines and two alternative embedding methods. For the feature baseline, we reuse the account-level features of~\cite{VALADARES2023120438}, namely transactions sent/received, transactions to/from smart contracts, Ether balance, and total transactions, and add API-derived features for transaction frequency, transaction volume, and token diversity. The second baseline is \method\!\!, a graph-based behavioural pipeline from our previous work~\citep{ZELENYANSZKI2026115102}. As alternative embedding methods, TF-IDF+SVD and Doc2Vec replace the two-step embedding process by producing text-based user representations from the users' sentences before clustering. Each representation is clustered with \textit{K}-means using the elbow-selected \(k\), and results are averaged over five random seeds. For TF-IDF+SVD and Doc2Vec, we use the same settings as the framework, namely (\texttt{max\_len}=\num{32}), \texttt{prefix} cropping, and \texttt{uniform} sampling, ensuring a fair comparison. The comparison set contains \num{10100} addresses from the first-month time window, consisting of pre-labelled addresses together with \num{10000} additional users, where each selected address has at least two behavioural sentences and at least one transaction with events, as \method\!\! operates on event-bearing transactions.

\begin{table*}[!t]
\centering
\caption{Comparison of our framework with baseline and alternative methods. \emph{SC, DBI, and CHI measure geometric clustering quality. \(\eta^2_{\ell}\), NMI\(_{\ell}\), and Gap measure length/activity leakage. SepScore, Sep-NMI, Purity, OneClus, Distinct, and Coll. measure pre-labelled category separation.}}
\label{tab:rq2-method-comparison}
\small
\setlength{\tabcolsep}{2pt}
\renewcommand{\arraystretch}{1.08}

\begin{tabular}{@{}lccccccccccccc@{}}
\hline
\textbf{Method} &
\(\boldsymbol{k}\) &
\textbf{SC} &
\textbf{DBI} &
\textbf{CHI} &
\(\boldsymbol{\eta^2_{\ell}}\) &
\(\mathbf{NMI}_{\ell}\) &
\textbf{Gap} &
\textbf{SepScore} &
\textbf{Sep-NMI} &
\textbf{Purity} &
\textbf{OneClus} &
\textbf{Distinct} &
\textbf{Coll.} \\
\hline

Feature &
\num{9} &
\num{0.412} &
\num{1.190} &
\num{3965.2} &
\num{0.701} &
\num{0.335} &
\num{0.517} &
\num{0.571} &
\num{0.147} &
\num{0.700} &
\num{0.280} &
\num{3.0} &
\num{0.0} \\

\method\!\! &
\num{6} &
\textbf{\num{0.691}} &
\textbf{\num{0.610}} &
\textbf{\num{60110.3}} &
\num{0.537} &
\num{0.268} &
\num{0.402} &
\num{0.341} &
\num{0.084} &
\num{0.650} &
\num{0.590} &
\num{1.0} &
\num{2.0} \\

TF-IDF+SVD &
\num{10} &
\num{0.192} &
\num{2.420} &
\num{597.4} &
\num{0.299} &
\num{0.151} &
\num{0.223} &
\num{0.593} &
\num{0.199} &
\num{0.752} &
\num{0.324} &
\num{3.0} &
\num{0.0} \\

Doc2Vec &
\num{9} &
\num{0.100} &
\num{2.891} &
\num{445.1} &
\num{0.291} &
\num{0.156} &
\num{0.223} &
\textbf{\num{0.621}} &
\num{0.237} &
\num{0.770} &
\num{0.274} &
\num{3.0} &
\num{0.0} \\

Our framework &
\num{8} &
\num{0.328} &
\num{1.409} &
\num{3048.6} &
\textbf{\num{0.248}} &
\textbf{\num{0.119}} &
\textbf{\num{0.182}} &
\num{0.603} &
\textbf{\num{0.240}} &
\textbf{\num{0.790}} &
\textbf{\num{0.260}} &
\num{2.6} &
\num{0.4} \\
\hline
\end{tabular}%
\end{table*}

Table~\ref{tab:rq2-method-comparison} summarises the comparison across conventional clustering quality, length/activity leakage, and pre-labelled category separation. The results should therefore be read across the three metric groups rather than as a single-score ranking. \method\!\! achieves the strongest geometric clustering results, with the highest SC and CHI and the lowest DBI, confirming its ability to form compact graph-based user groups. Its leakage values indicate that activity scale also contributes to the recovered structure, which is expected in DeFi user-flow data where interaction intensity and behavioural structure are closely related. The feature baseline shows even stronger leakage, indicating that its clusters are heavily influenced by account-level activity statistics. TF-IDF+SVD and Doc2Vec provide useful text-based alternatives, as both separate the pre-labelled categories into distinct best clusters; however, their low SC and high DBI indicate weaker overall cluster structure. In contrast, the proposed framework does not maximise the geometric clustering metrics, but it achieves the lowest length/activity leakage, the highest Sep-NMI, the highest labelled-cluster purity, and the smallest one-cluster malicious concentration. This makes it the only method that remains consistently strong across cluster geometry, leakage control, and pre-labelled category separation, rather than improving one dimension at the expense of another.

\begin{tcolorbox}[rqbox,title=Answer to RQ2]
The proposed framework provides the most suitable representation for behavioural user analysis. Although \method\!\! achieves stronger geometric scores, the proposed framework better balances cluster quality, leakage control, and separation.
\end{tcolorbox}
\section{Behavioural Analysis}
\label{sec:analysis}
This section addresses RQ3 and RQ4 by analysing the behaviours recovered from the selected clusters and their persistence over time. It first examines the first-month clusters to identify the main behavioural families, suspiciousness levels, and behavioural indicators. It then studies monthly and cumulative windows to assess whether ordinary and suspicious behavioural tags recur beyond a single observation period.

\subsection{RQ3: What Behavioural Characteristics Can Be Learned from This Type of Analysis?}
\label{sec:beh_analysis}
We interpret clusters using the behavioural profiler in Subsection~\ref{subsec:profiler}. The profiler summarises motifs, motif sequences, entity usage, temporal patterns, diversity, concentration, risk-label evidence, and risky-entity or exploit-transaction exposure. These summaries turn cluster memberships into behavioural indicators, explaining the roles, routines, and risk-associated patterns captured by the learned representations.


\subsubsection{Learned Behavioural Characteristics from the First-month Analysis}
\label{subsubsec:first_month_behavioural_characteristics}
The first-month clusters form a mixed behavioural landscape rather than being dominated by a single activity type. As summarised in Table~\ref{tab:first_month_behaviour_families}, the main families include transfer and approval behaviour, DEX/AMM interaction, NFT flows, mint/claim/reward activity, and ENS/name-service or bridge-like activity. These families describe behavioural settings rather than threat categories: the largest families involve ordinary stablecoin, transfer, approval, DEX/AMM, NFT, and reward-related behaviours, while suspicious evidence appears as smaller pockets within them. The behavioural reading combines motifs, ordered routines, temporal activity, diversity, concentration, direct labels, and risky-entity or exploit-transaction exposure into indicators that distinguish burst-like, persistent, broad, and entity-centred behaviour.

\begin{table}[!t]
\centering
\caption{First-month behavioural families.}
\label{tab:first_month_behaviour_families}
\small
\setlength{\tabcolsep}{2pt}
\renewcommand{\arraystretch}{1.05}
\begin{tabular}{@{}
>{\raggedright\arraybackslash}p{0.52\columnwidth}
>{\centering\arraybackslash}p{0.15\columnwidth}
>{\raggedright\arraybackslash}p{0.27\columnwidth}
@{}}
\hline
\textbf{Behaviour family} &
\textbf{Users} &
\textbf{Clusters} \\
\hline
Stablecoin / transfer / approval &
\num{1036073} &
1, 2, 6, 13, 14, 18 \\

DEX / AMM interaction &
\num{808653} &
3, 5, 16, 21 \\

NFT flow &
\num{527614} &
0, 4, 8, 10, 11, 20 \\

Mint / claim / reward &
\num{459932} &
7, 12, 15, 17 \\

ENS / name-service / bridge-like &
\num{199963} &
9, 19 \\
\hline
\end{tabular}
\end{table}

\begin{table}[!t]
\centering
\caption{First-month suspiciousness interpretation levels.}
\label{tab:first_month_suspiciousness_groups}
\small
\setlength{\tabcolsep}{2pt}
\renewcommand{\arraystretch}{1.05}
\begin{tabular}{@{}
>{\raggedright\arraybackslash}p{0.36\columnwidth}
>{\centering\arraybackslash}p{0.16\columnwidth}
>{\raggedright\arraybackslash}p{0.41\columnwidth}
@{}}
\hline
\textbf{Suspiciousness level} &
\textbf{Users} &
\textbf{Clusters} \\
\hline
Meaningfully suspicious &
\num{952868} &
5, 8, 13, 16, 3 \\

Watchlist &
\num{401962} &
0, 9, 18, 19, 21 \\

Mostly ordinary / exposure evidence &
\num{1677405} &
1, 2, 4, 6, 7, 10, 11, 12, 14, 15, 17, 20 \\
\hline
\end{tabular}
\end{table}

\paragraph{Stablecoin, Transfer, and Approval Behaviour}
Stablecoin and transfer-heavy clusters form the main ordinary baseline. Clusters 2 and 14 are the clearest stablecoin-transfer groups, while Clusters 1, 6, and 18 mix transfer behaviour with approval or exposure evidence. Cluster 13 is the main exception, combining approval management, longer activity histories, and the clearest phishing-focused signal in this family. The main indicators are repeated ERC20 transfers, stablecoin-specific movement, approval-linked transfer routines, and differences in activity span or token concentration.

\paragraph{DEX/AMM Behaviour}
DEX/AMM clusters capture swaps, routing, liquidity, pools, LP-token evidence, and repeated protocol interaction. Cluster 5 is the largest and strongest mixed-risk group, with bot, exploit, bug-exploit, phishing, and front-run evidence concentrated around a small set of entities. Cluster 3 shows active Uniswap-style trading with bot, exploit, and oracle manipulation pockets. Cluster 16 represents a broader DEX/AMM community, while Cluster 21 is best read as a DEX/approval watchlist community. These indicators distinguish ordinary trading communities from concentrated DEX groups where bot, exploit, front-run, or oracle manipulation evidence becomes behaviourally meaningful.

\paragraph{NFT Transfer and Approval Behaviour}
NFT clusters are characterised by ERC721/ERC1155 transfers, approval-for-all events, marketplace activity, minting, and repeated NFT-contract interaction. Cluster 8 is the clearest suspicious NFT cluster, where phishing and exploit evidence aligns with longer-lived NFT transfer and approval behaviour. Clusters 4, 10, and 20 provide weaker-risk NFT baselines, while Clusters 0 and 11 are better read as mixed or watchlist NFT/token-transfer communities. The key indicators are NFT transfer motifs, approval-for-all behaviour, marketplace-style interaction, minting, and repeated NFT-contract use.

\paragraph{Mint, Claim, and Reward Behaviour}
Mint, claim, and reward clusters describe campaign-like participation through claim, mint, reward, airdrop, and distribution motifs. Cluster 15 is the clearest burst-like campaign, while Clusters 12 and 17 show more persistent reward-style participation and Cluster 7 sits between burst-like and repeated reward behaviour. Risk exposure appears in this family, but the dominant behaviour remains campaign, claim, or reward participation. The key indicators are claim, mint, reward, airdrop, and distribution motifs, especially when they appear in repeated routines or are followed by token transfer.

\paragraph{ENS/name-service and Bridge-like Behaviour}
ENS/name-service and bridge-like clusters are smaller but distinct. Cluster 9 combines name-service management with NFT transfer and approval behaviour, while Cluster 19 combines name-service behaviour with bridge-like or deposit-like movement. Both are best read as specialised watchlist communities where risk evidence is embedded within coherent ENS/name-service or bridge-like behaviour. The indicators include name-service management, resolver or ownership-style updates, bridge-like or deposit-like movement, and repeated interaction with specialised identity or cross-chain entities.

The suspiciousness reading is grouped into three interpretation levels, as shown in Table~\ref{tab:first_month_suspiciousness_groups}. These are not binary normal/malicious labels, but indicate how strongly risk evidence shapes a cluster. Meaningfully suspicious clusters contain user-label or exposure evidence aligned with coherent behavioural indicators; watchlist clusters contain genuine but weaker or subtype-specific evidence; and ordinary/exposure clusters may touch risky entities or exploit-labelled transactions without being treated as cluster-wide suspicious.


The meaningfully suspicious clusters remain grounded in recognisable behaviour. Cluster 5 is a concentrated DEX/AMM community with bot, exploit, bug-exploit, phishing, and front-run evidence. Cluster 8 links NFT transfer and approval behaviour with phishing and exploit evidence, while Cluster 13 is the clearest phishing-focused case, combining approval management, phishing evidence, and asset-movement indicators. Clusters 3 and 16 are broader DEX/AMM communities where bot, exploit, oracle manipulation, exploit-transaction, and risky-entity evidence appears within trading and liquidity behaviour. In these cases, suspiciousness comes from the alignment between risk evidence and behaviour-specific indicators, not from labels alone. These examples show that the same risk label is interpreted differently depending on whether it appears in approval-heavy, NFT-centred, concentrated DEX, or broader trading behaviour. 

The watchlist clusters show visible but more cautious signals. Cluster 0 combines NFT and token-transfer behaviour with sparse bug-exploit and exploit evidence; Cluster 9 links ENS/name-service activity with exploit and oracle manipulation evidence; Cluster 18 is a stablecoin-transfer watchlist cluster with bot evidence; Cluster 19 combines ENS/name-service and bridge-like behaviour with bug-exploit, exploit, exploit-transaction, and risky-entity evidence; and Cluster 21 is a DEX/approval watchlist cluster with bug-exploit, exploit, exploit-transaction, and risky-entity exposure. Ordinary/exposure clusters provide behavioural baselines: they may contact risky contracts, risky tokens, malicious entities, or exploit-labelled transactions, but this contact is treated as exposure rather than proof of cluster-level suspiciousness.

At the tag level, the profiler summarises recurring behavioural and risk-related evidence across clusters. Approval-linked swap and transfer tags are indicated by approvals, swaps, routing, liquidity interaction, approval-for-all events, ERC20 or NFT transfers, and outgoing asset movement. DEX trading and liquidity tags are indicated by swaps, routing, pool or sync events, LP-token evidence, and repeated protocol use. NFT market and approval tags are indicated by ERC721/ERC1155 transfers, approval-for-all, minting, listing, sale activity, and repeated NFT-contract interaction. Mint/claim tags are indicated by claim, mint, reward, airdrop, and token-distribution routines, often followed by transfer.

Other tags describe broader behavioural structure, including repeated contract or protocol interaction, stablecoin-family movement, token and contract diversity, entropy, and top-entity concentration. These indicators distinguish broad, exploratory behaviour from protocol-centred, campaign-like, or concentrated behaviour. Suspicious tags are interpreted through alignment: phishing is tied to approval management, approval-for-all, NFT movement, asset outflow, and phishing-related entity contact; bot and front-running behaviour to repeated swaps, transfers, approvals, liquidity interaction, routing, regular activity, and concentrated protocol use; and exploit, bug-exploit, and oracle manipulation signals to exploit-transaction exposure, risky-entity exposure, repeated contract interaction, and trading, NFT, transfer, or protocol-interaction settings.


\begin{tcolorbox}[rqbox,title=Answer to RQ3,breakable]
The profiler interprets clusters through behavioural families such as transfers, DEX/AMM trading, NFT activity, mint/claim behaviour, and ENS or bridge-like interaction.

Suspiciousness is treated as variation within these settings, linking phishing, bot/front-running, and exploit-related tags to coherent behavioural evidence without treating exposure as proof of cluster-wide maliciousness.
\end{tcolorbox}

\begin{table*}[!t]
\centering
\caption{Suspiciousness tags and behavioural indicators.}
\label{tab:suspicious_tag_indicators}
\small
\setlength{\tabcolsep}{2pt}
\renewcommand{\arraystretch}{1.02}
\begin{tabular}{@{}
>{\raggedright\arraybackslash}p{0.16\textwidth}
>{\raggedright\arraybackslash}p{0.50\textwidth}
>{\raggedright\arraybackslash}p{0.27\textwidth}
@{}}
\hline
\textbf{Suspicious tag} &
\textbf{Behavioural indicators} &
\textbf{Reading} \\
\hline
Phishing &
Approvals, NFT movement, asset outflow &
Persistent approval/NFT risk. \\
Bot &
Repeated swaps/transfers, protocol concentration &
Persistent automation. \\
Exploit &
Risky/exploit exposure in DEX, NFT, ENS, or transfer behaviour &
Risk layer in ordinary behaviour. \\
Bug exploit &
Repeated protocol routines, exploit exposure &
Persistent; clearer over time. \\
Front-run &
DEX/AMM swaps, routing, pool interaction &
Persistent DEX subtype. \\
Oracle manipulation &
Swaps, pool/sync behaviour, liquidity use &
Clearer in cumulative windows. \\
Rug pull &
Token-centred activity, concentration, risky-token evidence &
Grows from second month. \\
Flash-loan attack &
DEX/AMM liquidity routines, exploit context &
Visible from 1--3 months. \\
Governance attack &
Governance/protocol routines, concentrated entity use &
Sparse longer-window signal. \\
MEV exploit &
DEX/AMM trading, routing, market-ordering context &
Sparse longer-window subtype. \\
Price manipulation &
DEX/AMM trading, pool/liquidity use &
Only clear in 1--6 months. \\
Replay attack &
Exploit context, repeated interaction patterns &
Only clear in 1--6 months. \\
\hline
\end{tabular}
\end{table*}

\subsection{RQ4: What Are the Persistent, Long-term Behavioural Patterns?}
\label{subsec:rq4_long_term_patterns}
The long-term analysis uses persistence as a validation signal. Here, \emph{long-term} means recurring beyond one monthly window, either in the independent second month or in the cumulative 1--2, 1--3, and 1--6 month windows. The second-month window checks whether tags reappear in an independent month, while the cumulative 1--2, 1--3, and 1--6 month windows show how behavioural evidence develops over longer horizons. The analysis is tag-level rather than cluster-level because cluster identifiers are window-specific, whereas behavioural and suspiciousness tags provide a comparable vocabulary across windows. For each tag, persistence is interpreted through recurring behavioural indicators, including motifs, temporal regularity, and risky-entity exposure. Together, each recurring tag and its supporting indicators define a long-term, persistent behavioural pattern. Persistence therefore strengthens interpretation only when the recurring tag is supported by recurring behavioural evidence.

Ordinary behavioural structure is stable across windows. Approval-linked transfer, approval-linked swap, DEX trading, liquidity provision, NFT market and approval behaviour, mint/claim campaigns, repeated contract or protocol interaction, and stablecoin-family movement all recur across independent and cumulative settings. Indicators include approvals, token or NFT movement, swaps, routing, pool or liquidity interaction, claim/mint routines, repeated contract touches, stablecoin token flows, and concentration or diversity patterns. Longer windows also make long-lived activity clearer, increasing from one tagged cluster in the first month to thirteen tagged clusters in the 1--6 month window. This shows that longer observation windows reveal changes in behavioural regularity and persistence, rather than just capturing more activity.

These persistent ordinary tags provide the behavioural background for suspiciousness interpretation. Approval-linked transfer and swap are common DeFi routines, but they become security-relevant when approval, transfer, asset-movement, or liquidity indicators align with phishing, exploit, bot, front-run, or oracle manipulation evidence. DEX trading is the main setting for several risk tags, while NFT market and approval behaviour is important for interpreting phishing and approval-related risk. Mint/claim campaigns and stablecoin-family movement remain mostly ordinary baselines unless stronger risk evidence aligns with their routines.

The suspiciousness tags show three broad persistence patterns. Some are visible early and persist, including phishing, bot, exploit, bug exploit, and front-run. Others become clearer in cumulative windows, especially oracle manipulation and rug pull. Sparse subtypes such as flash-loan attack, governance attack, MEV exploit, replay attack, and price manipulation mainly appear in longer windows. Table~\ref{tab:suspicious_tag_indicators} summarises the behavioural indicators used to interpret these tags.

Suspicious tags are interpreted through behavioural alignment rather than labels alone. Phishing remains tied to approval management, approval-for-all, NFT movement, and asset outflow. Bot and front-running behaviour are linked to repeated DEX/AMM swaps, routing, approvals, liquidity interaction, regular activity, and concentrated protocol use. Exploit and bug-exploit evidence appears as a risk layer inside ordinary DEX, NFT, ENS/name-service, transfer, and watchlist communities, especially where risky-entity or exploit-transaction exposure aligns with repeated interaction patterns. Longer windows connect scattered actions into recognisable routines and help distinguish broad ordinary communities from narrower concentrated risk pockets.

\begin{table}[!t]
\centering
\caption{Rarity distribution of behavioural indicators.}
\label{tab:indicator_rarity_summary}
\small
\setlength{\tabcolsep}{2pt}
\renewcommand{\arraystretch}{1.05}
\begin{tabular}{@{}
>{\raggedright\arraybackslash}p{0.18\columnwidth}
>{\raggedright\arraybackslash}p{0.38\columnwidth}
>{\centering\arraybackslash}p{0.20\columnwidth}
@{}}
\hline
\textbf{Rarity} &
\textbf{Definition} &
\textbf{Indicators} \\
\hline
Often &
Used by \num{5} or more tags &
\num{4} \\

Common &
Used by \num{3}--\num{4} tags &
\num{8} \\

Repeated &
Used by \num{2} tags &
\num{18} \\

Rare &
Used by \num{1} tag &
\num{76} \\
\hline
\end{tabular}
\end{table}

\subsubsection{Distribution of Behavioural Indicators}
\label{subsubsec:indicator_distribution}
The indicator distribution measures how often each behavioural indicator is used across the tag vocabulary. As shown in Table~\ref{tab:indicator_rarity_summary}, the distribution is strongly long-tailed: a small number of shared indicators form a DeFi and permission backbone, while most indicators remain tag-specific. Shared indicators include swaps, routing, pools, liquidity interaction, approvals, repeated protocol use, exploit or risk exposure, and concentration evidence. This suggests that the profiler is not simply detecting high activity or generic risk exposure, but separating distinct behavioural mechanisms. Approval is especially important because it connects ordinary DeFi and NFT activity with phishing, approval-heavy NFT behaviour, and exploit-related interpretations. However, approval is not suspicious by itself; it becomes meaningful only when aligned with asset movement, NFT transfer, phishing-related evidence, risky-entity exposure, or exploit-transaction exposure. Confidence therefore comes from several indicators aligning with the same behavioural reading, rather than from a single indicator.

\begin{tcolorbox}[rqbox,title=Answer to RQ4]
The long-term analysis shows persistent behavioural patterns across windows, including DEX/AMM activity, mint/claim campaigns, stablecoin movement, NFT approvals, and repeated protocol interaction. Suspiciousness is contextual rather than categorical: phishing is linked to approvals, NFT activity, and asset movement; bot/front-running to repeated DEX activity; oracle manipulation to swap, pool, liquidity, and protocol-interaction evidence; and rug pull to token-centred concentration and campaign-like transfers. Confidence increases when multiple indicators align with direct labels, risky-entity exposure, or exploit-transaction exposure.
\end{tcolorbox}
\section{Limitations}
\label{sec:limits}
This work has several limitations. The pre-labelled set is small because public reports cover only some attacks, bots, malicious contracts, exploiters, and exploited entities, so it is used as high-confidence validation evidence rather than exhaustive ground truth. Suspiciousness tags are cluster-level interpretive signals, not malicious/ benign labels for every user. The framework also depends on decoded events, address classification, token metadata, external risk references, and API-supported enrichment, so broader evaluation across chains, protocols, and periods is needed.

\section{Conclusion and future work}
\label{sec:conclusion}
This paper presented a scalable, application-agnostic framework for \emph{persistent behavioural pattern discovery} of Ethereum users. By transforming raw transactions into behavioural sentences, learning user-level sequence representations, clustering users, and profiling the resulting groups, the framework supports population-level analysis beyond transaction-centric or attack-specific views. The results show that the discovered groups capture both ordinary behaviour, such as stablecoin transfer, DEX trading, and risk-relevant variation, including phishing, oracle manipulation, and exploit-related evidence.

The cross-window analysis shows that these patterns persist beyond a single observation period. Recurring tags reappear across independent monthly and cumulative windows, while longer windows make sparse suspicious subtypes such as rug pull, flash-loan attack, governance attack, MEV exploit, replay attack, and price manipulation easier to interpret. Future work will extend the analysis across range of chains and applications, compare with classification-based and attack-specific baselines, and evaluate the learned representations on downstream risk classification and prediction.

\section*{Data availability}
The entire dataset and the code can be made available upon request.


\bibliographystyle{cas-model2-names}

\bibliography{references}



\end{document}